\documentclass[prb,twocolumn,showpacs]{revtex4-1}
\usepackage[utf8x]{inputenc}
\usepackage{amsbsy}
\usepackage{graphicx}
\usepackage{color}
\usepackage{dcolumn}
\usepackage{amsmath}
\usepackage{amssymb}
\usepackage{amsfonts}
\usepackage{cancel}
\usepackage[caption=false]{subfig}

\usepackage[colorlinks=true, citecolor=black, linkcolor=black, 
urlcolor=black]{hyperref}

\newcommand{\abinitio}{\emph{ab initio}}

\newcommand{\bra}[1]{\langle #1|}
\newcommand{\ket}[1]{|#1 \rangle }
\newcommand{\beq}{\begin{equation}}
\newcommand{\eeq}{\end{equation}}
\newcommand{\beqn}{\begin{eqnarray}}
\newcommand{\eeqn}{\end{eqnarray}}
\def\qq{\mathbf{q}}

\begin{document}
\title{First-principles calculations of phonon frequencies, lifetimes and spectral functions
       from weak to strong anharmonicity: the example of palladium hydrides}

\author{Lorenzo Paulatto$^{1}$}
\author{Ion Errea$^{2, 3}$}
\author{Matteo Calandra$^{1}$}
\author{Francesco Mauri$^{1}$}

\affiliation{$^1$Institut de min\'eralogie, de physique des mat\'eriaux et de 
cosmochimie (IMPMC), Universit\'e Pierre et Marie Curie (Paris VI), CNRS UMR 
7590, IRD UMR 206, Case 115, 4 place Jussieu, 75252 Paris Cedex 05, France}
\affiliation{$^2$Donostia International Physics Center (DIPC),
             Manuel de Lardizabal pasealekua 4, 20018 Donostia-San Sebasti\'an,
             Basque Country, Spain}
\affiliation{$^3$IKERBASQUE, Basque Foundation for Science, 48011, Bilbao, 
Spain}

\begin{abstract}
The variational stochastic self-consistent harmonic approximation
is combined with the calculation of third-order anharmonic 
coefficients within density-functional perturbation theory and the ``$2n+1$'' 
theorem to calculate
anharmonic properties of crystals. 
It is demonstrated that in the perturbative limit the combination
of these two methods yields the perturbative phonon linewidth and
frequency shift in a very efficient way, 
avoiding the explicit calculation of fourth-order anharmonic coefficients.
Moreover, it also allows calculating phonon lifetimes
and inelastic neutron scattering spectra in solids where the harmonic 
approximation breaks down and a non-perturbative
approach is required to deal with anharmonicity.
To validate our approach, we calculate the anharmonic phonon linewidth
in the strongly anharmonic palladium hydrides. 
We show that due to the large anharmonicity of hydrogen optical 
modes the inelastic neutron scattering spectra are not characterized by a 
Lorentzian line-shape,
but by a complex structure including satellite peaks.
\end{abstract}


\maketitle


\section{Introduction}
\label{introduction}

Anharmonic effects are responsible for the finiteness of the phonon
lifetimes observed experimentally as well as the temperature dependence
of the phonon frequencies. Indeed, harmonic phonon frequencies
are temperature independent and, as not decaying quasiparticles,
harmonic phonons' lifetime is infinite, yielding an artificial 
infinite value for the thermal conductivity.  
Considering that the calculation of the lattice thermal conductivity is crucial in 
semiconducting thermoelectrics~\cite{Mahan1997,DiSalvo30071999}
and in materials used for thermal dissipation~\cite{1.1524305},
the development of precise and reliable \abinitio{} approaches
based on density-functional theory (DFT)
to calculate anharmonic effects, including the calculation of lifetimes,
is currently a strongly active field of research.

\begin{figure}[t]
\includegraphics[width=1.0\linewidth]{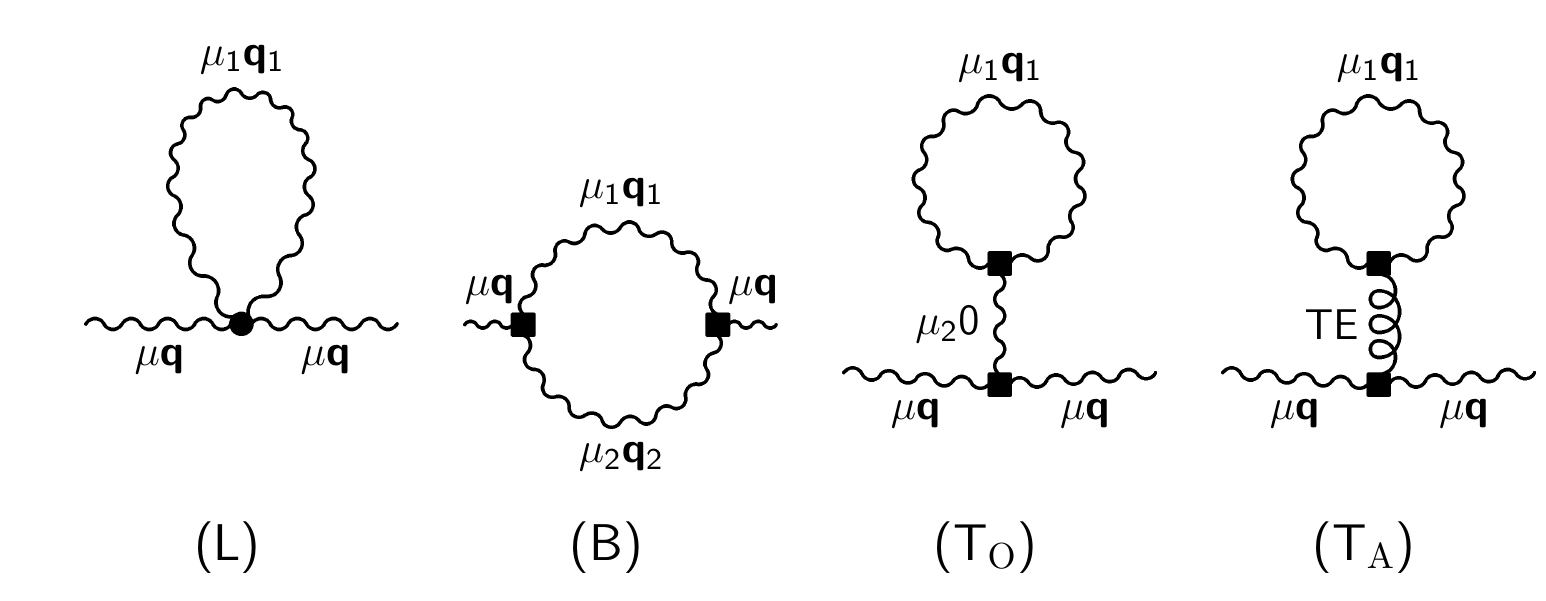}
\caption{The loop (L), bubble (B) and tadpole (T) diagrams 
contribute to the phonon self-energy
$\Pi_{\mu}(\qq,\omega)$ at lowest order
in perturbation theory.
The tadpole diagram is split into the optical (T$_\mathrm{O}$)
and acoustic (T$_\mathrm{A}$) contributions. The latter
accounts for the thermal expansion (TE). 
The dot denotes a fourth-order vertex, the
square a third-order one.}
\label{feynman-anharmonicity}
\end{figure}

Anharmonic effects can be treated 
perturbatively as long as they are weak. At lowest order, the harmonic phonon self-energy
should be corrected by the four Feynman self-energy diagrams
represented in 
Fig.~\ref{feynman-anharmonicity}~\cite{Maradudin62,Calandra07}.
The real part of the diagrams contributes to the phonon lineshift, i.e., to a change 
of the phonon frequency, while the imaginary part of the diagrams is responsible 
for the broadening of phonon lines. 
The three phonon process described by the bubble diagram 
contributes to both the linewidth and the lineshift, while the loop diagram 
has a two-phonon fourth-order vertex which contributes exclusively to the 
lineshift. We have finally decomposed the tadpole diagram in an optical 
part (T$_\mathrm{O}$), which contributes to the relaxation of internal coordinates, and an 
acoustical part (T$_\mathrm{A}$), where the line of zero energy and momentum can be associated 
to the elastic constants of the material and accounts for the thermal expansion. 
The tadpole diagrams reduce to the quasi-harmonic approximation (QHA) 
under certain conditions~\cite{Lazzeri98}.

The summation of these diagrams \emph{ab initio} requires the knowledge of 
third- and fourth-order derivatives of  the Born-Oppenheimer (BO) total energy. 
During the last 20 years several 
implementations have emerged to efficiently compute the third-order derivatives via 
the ``$2n+1$'' theorem~\cite{Gonze89,Debernardi95,Lazzeri02,Deinzer03,Paulatto13}.
On the contrary, the fourth order term cannot be computed in
this framework, but it can be obtained by finite difference calculations
of the third-order diagrams. 
As the number of fourth-order derivatives quickly increases 
with the number of atoms, the perturbative evaluation of the loop
diagram is only affordable for few atoms highly symmetric cells~\cite{Bonini07}.
Thus, new approaches to calculate fourth-order terms from first-principles 
are highly desirable.
Despite such difficulties, the perturbative approach has been successfully exploited 
in many different applications, ranging from computing phonon 
lifetimes~\cite{Lazzeri98,Shukla03} and mean free paths~\cite{Paulatto13}, 
to thermal expansion~\cite{Lazzeri02} and thermal conductivity of materials in 
the single mode approximation~\cite{Broido04,*Broido05,*Ward09,Paulatto13} or solving 
exactly the Boltzmann transport equation~\cite{Fugallo13}.

On the other hand, the perturbative approach to anharmonicity is built on 
top of a harmonic description of the material and it breaks down when the 
harmonic model does. For instance when a material exhibits imaginary phonon 
frequencies, either because of a real mechanical instability or because the 
phase is stabilized by quantum or thermal fluctuations strongly affected by the
anharmonic potential. This is the case in many 
ferroelectrics~\cite{PhysRevB.52.6301,PhysRevLett.74.4067} and
thermoelectrics~\cite{delaire614,PhysRevLett.112.175501}. 
In systems with light atomic species, the large amplitude rattling 
vibrations can also make the harmonic approximation
break down~\cite{PhysRevB.82.104504,PhysRevLett.110.025903}. Even the presence of 
soft modes associated with charge-density waves~\cite{PhysRevB.86.155125,
PhysRevB.80.241108,PhysRevLett.106.196406,PhysRevB.77.165135,PhysRevB.73.205102} 
can invalidate assumptions 
of perturbative methods. Calculating phonon lifetimes and, consequently, 
the thermal conductivity in all these systems requires, therefore, 
going beyond the perturbative approach.

In this work we want to tackle these problems by combining the 
third-order perturbative approach, which provides the leading order contribution 
to the phonon line broadening, with the stochastic self-sonsistent
harmonic approximation (SSCHA) method~\cite{Errea13,Errea14}, which 
is based on a variational approach and gives an accurate 
description of the phonon spectra in the non-perturbative regime. 
We show that for small anharmonicity the combination of these methods
accounts for all the diagrams in Fig. \ref{feynman-anharmonicity}
yielding the correct perturbative limit. Thus, the perturbative
correction can be obtained avoiding the cumbersome
and expensive calculation of fourth-order force-constants. In fact, in
this limit{\bf ,} the
SSCHA accounts for T$_\mathrm{O}$ and T$_\mathrm{A}$ tadpole diagrams
as well as the loop diagram, while the bubble diagram is computed 
making use of the third-order force-constants calculated with 
the ``$2n+1$'' theorem over the whole Brillouin Zone (BZ)~\cite{Paulatto13}.
In the non-perturbative regime, combining the
SSCHA equilibrium positions, phonon frequencies and polarization vectors with the calculated
third-order coefficients, phonon lifetimes and spectral functions
can be obtained for strongly anharmonic crystals where the harmonic
approximation breaks down.

We apply this method to palladium hydride stoichiometric compounds: 
PdH, PdD and PdT. Because of the huge anharmonicity of hydrogen rattling 
vibrations the phonon dispersions predicted by density-functional perturbation 
theory (DFPT) show strong instabilities and a huge underestimation of
the H-character optical modes.
The instabilities are even more pronounced at finite temperature when the
thermal expansion is considered. These instabilities are however not physical and can be 
cured with the SSCHA approach, which yields phonon spectra in good agreement with 
experiments~\cite{Errea13,Errea14}.
The anharmonicity also causes an interesting negative superconducting isotope 
coefficient~\cite{Errea13}. Here we find that, despite being a metal,
the phonon linewidth is dominated by the  phonon-phonon interaction 
for any temperature above a few Kelvin. 
The third-order broadening of the phonon modes is huge, 
at the point that the simplistic model of well-distinct phonon modes, broadened 
to a finite-width Lorentzian function, cannot be applied anymore.
It is remarkable, however, that the phonon shift induced by the 
third-order is not predominant 
over the SSCHA correction of the phonon bands. 


\section{The ionic Hamiltonian: phonon anharmonicity and the electron-phonon 
coupling}
\label{ionic-h}

Within the BO approximation, the dynamics of the ions in a crystalline solid
are determined by the following Hamiltonian:
\beq
H = T + V,
\label{h-total}
\eeq
where
\beq
T = \sum_{s \alpha \mathbf{R}} \frac{
            ( P^{\alpha}_s (\mathbf{R}))^2 }{2 M_s}
\label{kinetic}
\eeq
is the kinetic-energy operator of the ions and
$V$ the potential defined by the BO energy surface.
Let $s$ denote an ion within the unit cell,
$\mathbf{R}$ a lattice vector and $\alpha$ a Cartesian 
direction. 
In the equation above $P^{s \alpha}(\mathbf{R})$ is the momentum operator 
and $M_s$ the mass of ion $s$.
Considering that generally ions vibrate around their 
equilibrium position ${\bf R}_{{\rm eq}}^{s \alpha}$ determined
by the minima of the BO energy surface, the potential $V$ 
is Taylor-expanded as a function of the
ionic displacements from equilibrium 
$u_s^{\alpha}(\mathbf{R})$ as
\beq
V = V_0 + \sum_{n=2}^{\infty} V_n ,
\label{potential-taylor}
\eeq
where
\beq
V_n = \frac{1}{n!} \sum_{\substack{s_1 \dots s_n \\ 
\alpha_1 \dots \alpha_n \\ \mathbf{R}_1 \dots \mathbf{R}_n}}
    \phi_{s_1 \dots s_n}^{\alpha_1 \dots \alpha_n} (\mathbf{R}_1, \dots, 
\mathbf{R}_n) u_{s_1}^{\alpha_1}(\mathbf{R}_1) \dots  
u_{s_n}^{\alpha_n}(\mathbf{R}_n)
\label{potential-taylor-terms}
\eeq 
and
\beq
\phi_{s_1 \dots s_n}^{\alpha_1 \dots \alpha_n} (\mathbf{R}_1, \dots, 
\mathbf{R}_n) = \left[ \frac{ \partial^{(n)}
  V }{ \partial u_{s_1}^{\alpha_1}(\mathbf{R}_1) \dots \partial 
u_{s_n}^{\alpha_n}(\mathbf{R}_n)}
  \right]_0
\label{derivative-potential}
\eeq
represents the $n$-th order derivative of the BO energy surface
with respect to the atomic displacements calculated at
equilibrium, namely, the $n$-th order, or $n$-bodies, force-constants.

The Hamiltonian in Eq.~\eqref{h-total} represents a complicated 
many-body problem, unsolvable unless an approximated scheme
is adopted. 

\subsection{The harmonic approximation}
\label{harmonic-app}

The first non-trivial approximation is the harmonic
approximation, in which the expansion in Eq.~\eqref{potential-taylor}
is truncated at second order. In Fourier
space the Hamiltonian looks like
\beq
H_2 = \sum_{s \alpha \qq} \frac{
            |P_{s}^{\alpha}(\qq)|^2 }{2 M_s}
            + \frac{1}{2} \sum_{\substack{s_1 s_2 \\ \alpha_1 \alpha_2 \\ 
\qq}}
               u_{s_1}^{\alpha_1 *}(\qq) \phi_{s_1 s_2}^{\alpha_1 
\alpha_2} (\qq)
               u_{s_2}^{\alpha_2}(\qq) .
\label{harmonic-potential}
\eeq
If the phonon frequencies $\omega_{\mu} (\qq)$ and polarizations
$\epsilon_{\mu s}^{\alpha} (\qq)$ diagonalize the dynamical
matrix at momentum $\qq$,
\beq
\sum_{s_2 \alpha_2} \frac{\phi_{s_1s_2}^{\alpha_1 
\alpha_2}(\qq)}{\sqrt{M_{s_1}M_{s_2}}}
             \epsilon_{\mu s_2}^{\alpha_2} (\qq) =
             \omega^2_{\mu} (\qq) \epsilon_{\mu s_1}^{\alpha_1} 
(\qq),
\label{eigenvalue-eq}
\eeq
by applying the following change of variables
\beqn
u_{s}^{\alpha}(\qq) & = & \sum_{\mu} \sqrt{\frac{\hbar}{2 M_s 
\omega_{\mu} (\qq)}}
                                  \epsilon_{\mu s}^{\alpha} (\qq) 
[b_{\mu}(\qq) + b^{\dagger}_{\mu}(-\qq)]
                         \label{qmu} \\
P_{s}^{\alpha}(\qq) & = & \frac{1}{i} \sum_{\mu}\! \sqrt{\frac{\hbar M_s 
\omega_{\mu} (\qq)}{2}}
                                  \epsilon_{\mu s}^{\alpha} (\qq) 
[b_{\mu}(\qq) - b^{\dagger}_{\mu}(-\qq)]
                         \label{pmu}
\eeqn
$H_2$ can be written in this well-known diagonal form:
\beq
H_2 = \sum_{\qq \mu} \hbar \omega_{\mu} (\qq) 
       \left(b^{\dagger}_{\mu}(\qq) b_{\mu}(\qq)  + \frac{1}{2} 
\right)\,.
\eeq 
Here $b^{\dagger}_{\mu}(\qq)$ and $b_{\mu}(\qq)$ are, 
respectively, 
the phonon creation and annihilation operators. 
Thus, if the ionic Hamiltonian is approximated by $H_2$, phonons
are well-defined quasiparticles that do not interact nor decay.

\subsection{Anharmonic effects and perturbation theory}
\label{perturbation-theory}

It is convenient to write the Hamiltonian $H$ in Fourier space
as~\cite{Calandra07}
\beqn
H & = & H_2 + \sum_{n=3}^{\infty} \frac{N^{1-n/2}}{n!}  
    \sum_{\mu_1 \dots \mu_n \atop \qq_1 \dots \qq_n}
    \phi_{\mu_1 \dots \mu_n}(\qq_1, \dots , \qq_n) \nonumber \\ 
& \times &  A_{\mu_1}(\qq_1) \dots A_{\mu_n}(\qq_n),
\label{potential-taylor-mode}
\eeqn
where $A_{\mu}(\qq) = b_{\mu}(\qq) + 
b^{\dagger}_{\mu}(-\qq)$.
The reader is referred to Appendix \ref{appen-fourier}
for the definition of the high-order $\phi_{\mu_1 \dots \mu_n}(\qq_1, 
\dots , \qq_n)$
anharmonic coefficients. Due to translational symmetry, for each  
$\phi_{\mu_1 \dots \mu_n}(\qq_1, \dots , \qq_n)$ coefficient
crystal momentum is conserved so that
$\qq_1 + \dots + \qq_n = \mathbf{G}$, where $\mathbf{G}$
is a reciprocal lattice vector~\cite{PhysRevB.82.104504}. 

Within perturbation theory, the lowest energy 
corrections to the harmonic non-interacting 
phonon propagator are given by the
loop (L), bubble (B) and tadpole (T) self-energy diagrams
shown in Fig.~\ref{feynman-anharmonicity}.
Their contribution to the self-energy are given 
by~\cite{Maradudin62,PhysRevB.82.104504,Calandra07,PhysRevB.68.220509}
\begin{align}
\Pi^{L}_{\mu }(\qq,\omega)  &=
  \frac{1}{2N} \sum_{\qq_1 \mu_1} 
    \phi_{\mu \mu \mu_1\mu_1}(-\qq,\qq, \qq_1, -\qq_1) 
\nonumber
\\
  &\times  
  [2n_B(\omega_{\mu_1}(\qq_1)) + 1],
\label{pi-l}
\\
\Pi^{B}_{\mu }(\qq,\omega) &= - \frac{1}{2N} 
  \sum_{\substack{\qq_1 \qq_2 \\ \mu_1 \mu_2}} \sum_{\mathbf{G}}
  \delta_{-\qq + \qq_1 + \qq_2, \mathbf{G}}
\nonumber
\\ 
&\times  
  |\phi_{\mu \mu_1\mu_2}(-\qq, \qq_1, \qq_2)|^2 F(\omega,\omega_{\mu_1}(\qq_1),\omega_{\mu_2}(\qq_2)), 
\label{pi-b}
\\ 
\Pi^{T}_{\mu }(\qq,\omega) &= - \frac{1}{N} 
  \sum_{\substack{\qq_1 \\ \mu_1 \mu_2}} 
  \phi_{\mu \mu \mu_2}(-\qq,\qq, 0)
  \phi_{\mu_1 \mu_1\mu_2}(-\qq_1,\qq_1, 0)
\nonumber
\\
  &\times 
  \frac{2n_B(\omega_{\mu_1}(\qq_1))+ 1}{\hbar \omega_{\mu_2}(0) },
\label{pi-t}
\end{align}
where
\begin{align}
F(\omega,\omega_1,\omega_2) & = \frac{1}{\hbar}  \Bigg[
                                       \frac{2(\omega_1 + \omega_2)
                                             [1 + n_B(\omega_1) + 
n_B(\omega_2)]}{
                                              (\omega_1 + \omega_2)^2 - 
                                              (\omega + i \delta)^2} \nonumber 
\\
                                      & +
                                       \frac{2(\omega_1 - \omega_2)
                                             [n_B(\omega_2) - n_B(\omega_1)]}{
                                              (\omega_2 - \omega_1)^2 - 
                                              (\omega + i \delta)^2}
                                       \Bigg],
\label{f-fun}
\end{align}
$n_B(\omega) = 1 / \left( e^{\beta \hbar \omega} - 1\right)$
is the bosonic occupation factor and $N$ the number of unit cells in the system.
Thus,
\begin{align}
\Pi_{\mu}(\qq,\omega) = \Pi^{L}_{\mu }(\qq,\omega) +
                                    \Pi^{B}_{\mu }(\qq,\omega) +
                                    \Pi^{T}_{\mu }(\qq,\omega) 
\end{align}
is the total self-energy at the perturbative level.

The tadpole T$_\mathrm{O}$ diagram includes an optical phonon at the 
$\Gamma$ point ($\omega_{\mu_2}(0)$) and it accounts for the relaxation 
of internal coordinates. If internal coordinates are determined 
by symmetry, it vanishes and does not 
contribute~\cite{Maradudin62,Calandra07,PhysRevB.68.220509}.
On the contrary, if Wyckoff's positions have free parameters,
the T$_\mathrm{O}$ diagram accounts for the effect of quantum
and thermal fluctuations in the internal coordinates. 
Its effect can be accounted if internal coordinates are those 
that minimize the QHA free energy. 
For those internal coordinates, the T$_\mathrm{O}$ self-energy vanishes.
These equilibrium positions 
might differ from the $R_{{\rm eq}}^{s \alpha}$ positions 
derived from the minima of the BO energy surface.
The related T$_\mathrm{A}$ diagram includes the limit for 
${\bf q} \rightarrow \Gamma$ of an acoustical phonon, 
and accounts for the relaxation of the cell parameters. 
It can also be calculated within the QHA~\cite{Bonini07}.

The bubble contribution is the only self-energy term with
an imaginary part. Thus, it is the only one contributing to the
phonon linewidth. 
As long as $|\Pi_{\mu}(\qq,\omega)| \ll \hbar \omega_{\mu} (\qq)$,
the half-width at half-maximum (HWHM) of phonon $\mu$ with momentum
$\qq$ is given by the imaginary part of
the bubble self-energy term at the harmonic frequency, namely,
\beq
\Gamma^{\mathrm{ph-ph}}_{\mu}(\qq) = - \mathfrak{Im} 
                      \Pi^{B}_{\mu}(\qq,\omega_{\mu} (\qq)).
\label{hfhm}
\eeq
The shift of the harmonic frequency due to the phonon-phonon
interaction is given instead by the real part of both loop
and bubble diagrams,
\beq
\Delta_{\mu}(\qq) = \mathfrak{Re} \left[ 
                      \Pi^{L}_{\mu}(\qq,\omega_{\mu} (\qq))
                + \Pi^{B}_{\mu}(\qq,\omega_{\mu} (\qq)) 
\right].
\label{shift}
\eeq
The frequency shifted by anharmonicity is thus
$\omega_{\mu} (\qq) + \Delta_{\mu}(\qq)$.

\subsection{The non-perturbative limit and the stochastic self-consistent
            harmonic approximation}
\label{scha}

The perturbative scheme introduced in 
Sec.~\ref{perturbation-theory} assumes that phonons 
can be described as well-distinct quasi-particle states. 
This assumption is only valid when the anharmonic self-energy 
is small with respect to the separation between harmonic frequencies. 
If $|\Pi_{\mu}(\qq,\omega)|$
is comparable to or larger than $\hbar \omega_{\mu} (\qq)$, it breaks 
down and the phonon linewidhts and shifts derived from Eqs. \eqref{hfhm}
and \eqref{shift} are not correct. In case the system has
harmonic imaginary frequencies, i.e., it is unstable in the harmonic
approximation, the perturbative approach cannot even be applied. 

A useful workaround that allows treating this non-perturbative regime
is the variational approach defined by the
self-consistent harmonic approximation (SCHA)~\cite{hooton422},
which has been recently implemented in a stochastic framework within the
SSCHA~\cite{Errea13,Errea14}.
The exact vibrational free energy of a solid is given by
the sum of the total energy and the entropy:
\beq
F_H = \mathrm{tr} \left( \rho_H H \right) + \frac{1}{\beta} 
      \mathrm{tr} \left( \rho_H \ln \rho_H \right),
\label{true-free-energy}
\eeq 
where the $\rho_H = e^{- \beta H} / [\mathrm{tr} ( e^{- \beta H})]$ is the 
density
matrix and $\beta = 1 / (k_BT)$. 
In the SSCHA 
$\rho_H$ is substituted by a trial density matrix $\rho_{\mathcal{H}}$ 
defined by a trial 
$\mathcal{H} = T + \mathcal{V}$ Hamiltonian. 
Then, if
\beq
\mathcal{F}_H[\mathcal{H}] = \mathrm{tr} \left( \rho_{\mathcal{H}} H \right) + 
      \frac{1}{\beta} 
      \mathrm{tr} \left( \rho_{\mathcal{H}} \ln \rho_{\mathcal{H}} \right)
\label{var-free-energy}
\eeq
we have the 
\beq
F_H \le \mathcal{F}_H[\mathcal{H}] = F_{\mathcal{H}} + \mathrm{tr} \left[ 
\rho_{\mathcal{H}}
      ( V - \mathcal{V}) \right]
\label{gibbs-bogoliubov}
\eeq  
variational equation. Thus, minimizing $\mathcal{F}_H[\mathcal{H}]$
with respect to the trial $\mathcal{H}$ Hamiltonian a good 
approximation of the free energy is obtained.

Even if the variational principle in Eq.~\eqref{gibbs-bogoliubov}
is valid for any trial potential $\mathcal{V}$, in the SSCHA
we assume that $\mathcal{V}$ is a harmonic potential
that can be parametrized as
\beq
\mathcal{V} = \frac{1}{2} \sum_{\qq}
              \sum_{\substack{s_1 s_2 \\ \alpha_1 \alpha_2 }}
               \mathfrak{u}_{s_1}^{\alpha_1 *}(\qq) \Phi_{s_1 
s_2}^{\alpha_1 \alpha_2} (\qq)
               \mathfrak{u}_{s_2}^{\alpha_2}(\qq).
\label{potential}
\eeq
The $\mathfrak{u}_{s}^{\alpha}(\qq)$ atomic displacements
are different from the $u_{s}^{\alpha}(\qq)$ displacements.
The latter measure the displacement of the atoms from $R_{{\rm eq}}^{s 
\alpha}$, 
while the first represent the displacement from
trial equilibrium positions, $\mathfrak{R}_{{\rm eq}}^{s \alpha}$. 
Similarly the $\Phi (\qq)$ force-constants
are not the harmonic ones of Eq.~\eqref{harmonic-potential},
but some trial force-constants.
Thus, the independent parameters in $\mathcal{V}$
are simply $\mathfrak{R}_{{\rm eq}}^{s \alpha}$ and $\Phi (\qq)$.
In the SSCHA $\mathcal{F}_H[\mathcal{H}]$ is minimized with respect to
these parameters making use of a conjugate-gradient
(CG) algorithm. During the minimization crystal symmetries are preserved. 
At the minimum, the $\mathfrak{R}_{{\rm eq}}$ positions 
are the equilibrium positions including anharmonic effects 
and the phonon frequencies and polarizations
obtained diagonalizing $\Phi(\qq)$
yield the phonon spectra renormalized by anharmonicity.    

The gradient of $\mathcal{F}_H[\mathcal{H}]$
needed for the CG minimization is given 
in Fourier space by~\cite{Errea13,Errea14}
\beqn
&& 
\boldsymbol{\nabla}_{\mathbf{\mathfrak{R}}_{\mathrm{eq}}}\mathcal{F}_H[\mathcal{
H}] =
  - \langle \mathbf{f}(0) - \boldsymbol{\mathfrak{f}}(0) \rangle_{\mathcal{H}}
\label{gradient-eq} \\
&& \boldsymbol{\nabla}_{\Phi(\qq)} \mathcal{F}_H[\mathcal{H}] =   
   -\sum_{ss'\alpha \alpha' \mu} 
\sqrt{\frac{M_{s'}}{M_s}} 
    \Big[ \varepsilon_{\mu s}^{\alpha} (\qq)  
     \boldsymbol{\nabla}_{\Phi(\qq)} \ln \mathfrak{a}_{\mu}(\qq)
\nonumber \\ && \ \ 
   + \boldsymbol{\nabla}_{\Phi(\qq)} \varepsilon_{\mu s}^{\alpha} 
(\qq) 
    \Big] \varepsilon_{\mu s'}^{\alpha' *} (\qq)  
   \langle [ f_s^{\alpha}(\qq) - \mathfrak{f}_s^{\alpha}(\qq) ]^* 
\mathfrak{u}_{s'}^{\alpha'}(\qq)
   \rangle_{\mathcal{H}}, 
\label{gradient-phi}
\eeqn
where $\langle O \rangle_{\mathcal{H}} = \mathrm{tr} ( \rho_{\mathcal{H}} O )$ 
represents
the quantum statistical average calculated with $\mathcal{H}$.
In Eqs. \eqref{gradient-eq} and \eqref{gradient-phi}
$\mathbf{f}(\qq)$ is the vector formed
by all the atomic forces  at momentum $\qq$ and
$\boldsymbol{\mathfrak{f}}(\qq)$
denotes the vector formed by the forces derived from $\mathcal{V}$.
The normal length is
$\mathfrak{a}_{\mu}(\qq)   =\sqrt{\hbar \coth (\beta \hbar \Omega_{\mu} 
(\qq) / 2) / 
                                         (2 \Omega_{\mu} (\qq) )}$,
and $\Omega_{\mu} (\qq)$ and $\varepsilon_{\mu s}^{\alpha} (\qq)$
are different from the harmonic phonon frequencies and polarizations
since they are obtained diagonalizing $\Phi(\qq)$
and not $\phi(\qq)$.
In the SSCHA the quantum statistical averages in Eqs.
\eqref{gradient-eq} and \eqref{gradient-phi}
are calculated stochastically calculating atomic forces
on supercells with ionic configurations described
by $\rho_{\mathcal{H}}(\mathbf{R}) = 
\bra{\mathbf{R}} \rho_{\mathcal{H}} \ket{\mathbf{R}}$,
where $\mathbf{R}$ represents a general ionic configuration~\cite{Errea13,Errea14}.

\subsection{Combining the SSCHA with the perturbative expansion}
\label{scha-bubble}

The internal coordinates determined by the SSCHA are those
that make the quantum statistical average of the forces
on the ions vanish. As shown in Appendix~\ref{appen-scha},
in the perturbative limit this means that the T$_\mathrm{O}$
diagram vanishes for those coordinates. The reason is that
the tadpole diagram includes the $\sum_{\qq_1 \mu_1} \phi_{\mu_1 \mu_1\mu_2}(-\qq_1,\qq_1, 0)$
term, which vanishes if the quantum statistical average of the forces is zero.
The T$_\mathrm{A}$ diagram that describes the thermal expansion is also
included in the SSCHA if the lattice parameters are determined by the
minima of the free energy~\cite{Errea14}. Therefore, in the perturbative
limit the SSCHA reduces to the QHA. 

On the other hand, we also show in Appendix \ref{appen-scha}
that for a fixed cell geometry the perturbative limit
of the frequency shift given by the SSCHA is
exactly the shift given by the loop self-energy
term. 
Therefore, the SSCHA provides a stochastic framework to 
calculate the fourth-order anharmonic shift in systems were perturbation theory works.
This is very remarkable as with the SSCHA we avoid the cumbersome and time-demanding calculation of
$\phi_{\mu \mu \mu_1\mu_1}(-\qq, \qq, \qq_1, -\qq_1)$
anharmonic coefficients needed in Eq.~\eqref{pi-l}, which are usually calculated
taking first-order numerical derivatives of third-order anharmonic terms
or second-order numerical derivatives of dynamical matrices
calculated in supercells~\cite{PhysRevB.59.6182,PhysRevB.68.220509,PhysRevB.82.104504,
PhysRevLett.106.165501,errea:112604,Bonini07}.
This result also indicates that in the perturbative limit the
SSCHA does not include the bubble self-energy. Notwithstanding that, 
it shows us that the bubble self-energy can be included
on top of the SSCHA result.

Once the minimization procedure of the SSCHA has been performed, 
the difference between the SSCHA and the exact potential is 
lead by the third-order term:
$V_3 \sim V-\mathcal{V}$.
The lowest-energy
third-order contribution to the self-energy is
obtained by substituting the SSCHA equilibrium positions,
phonon frequencies and polarizations in Eq.~\eqref{pi-b},
\beqn
&& \Pi^{\mathcal{H}B}_{\mu }(\qq,\omega)  =  - \frac{1}{2N} 
  \sum_{\substack{\qq_1 \qq_2 \\ \mu_1 \mu_2}} \sum_{\mathbf{G}}
  \delta_{-\qq + \qq_1 + \qq_2, \mathbf{G}}
\nonumber
\\ 
&& \ \ \ \times  
  |\phi^{\mathcal{H}}_{\mu \mu_1\mu_2}(-\qq, \qq_1, \qq_2)|^2 F(\omega,\Omega_{\mu_1}(\qq_1),
  \Omega_{\mu_2}(\qq_2)), 
\label{pi-b-sscha}
\eeqn
where the $\mathcal{H}$ superscript denotes that the anharmonic coefficients 
and the self-energy are calculated with the SSCHA equilibrium positions, 
phonon frequencies and
polarizations (see Appendix~\ref{appen-fourier}). 
Then, the linewidth is
$\Gamma^{\mathrm{ph-ph}}_{\mu}(\qq) = - \mathfrak{Im} 
\Pi^{\mathcal{H}B}_{\mu}(\qq,\Omega_{\mu} (\qq))$
and the SSCHA phonon frequencies can be
corrected with the 
$\Delta_{\mu}(\qq) = \mathfrak{Re} 
 \Pi^{\mathcal{H}B}_{\mu}(\qq,\Omega_{\mu} (\qq))$ shift.
In the perturbative limit, when $\mathcal{V} = V_2$, the bubble self-energy
in Eq. \eqref{pi-b-sscha} trivially reduces to the perturbative one in Eq.~\eqref{pi-b}.
As the SSCHA includes both the tadpole and loop self-energy corrections, 
including this third-order contribution yields the correct perturbative
limit. 
In the non-perturbative limit on the contrary, 
we assume that the lowest-energy correction to the SSCHA
result is dominated by Eq. \eqref{pi-b-sscha}. 
This type of ``bubble'' correction to the SCHA phonon frequencies
was already applied by Cowley using a model potential
for a $\Gamma$ point mode of a diatomic ferroelectric~\cite{Cowley1996585}.
This framework, therefore, allows us to calculate
not only bubble self-energy corrections to the SSCHA
result, but to calculate anharmonic lifetimes in strongly
anharmonic systems, opening the door to the calculation
of the thermal conductivity in crystals were the perturbative
expansion breaks down.

\subsection{The electron-phonon linewidth}
\label{electron-phonon}

For metals the phonons acquire an additional linewidth contribution
due to the electron-phonon interaction.
In complete analogy with the anharmonic case,
the phonon linewidth associated to the
electron-phonon interaction is given by the imaginary
part of the phonon self-energy associated to the electron-phonon
interaction. Within Migdal's lowest-order approximation
for the self-energy, 
the linewidth at HWHM reads~\cite{allen-mitrovic,PhysRevB.71.064501}
\beqn
\Gamma^{\mathrm{el-ph}}_{\mu}(\qq) & = & \frac{2\pi}{N} \sum_{\mathbf{k} 
m n}
                      |g^{\mu}_{\mathbf{k}n,\mathbf{k}+\qq m}|^2
                      (f_{\mathbf{k}n} - f_{\mathbf{k}+\qq m}) \nonumber 
\\ & \times &  
                      \delta(\epsilon_{\mathbf{k}n} - 
\epsilon_{\mathbf{k}+\qq m} + \hbar \omega_{\mu}(\qq)). 
\label{hfhm-elph}
\eeqn
In Eq.~\eqref{hfhm-elph}, $g^{\mu}_{\mathbf{k}n,\mathbf{k}+\qq m} = 
\bra{{\bf k}n}
\delta V / \delta q_{\mu}(\qq) \ket{{\bf k}+{\bf q}m} \sqrt{\hbar / (2 
\omega_{\mu}({\bf q}))}$ 
is the electron-phonon matrix element,
$\ket{{\bf k}n}$ a
Kohn-Sham state with energy $\epsilon_{{\bf k}n}$ measured from the Fermi 
energy,
$q_{\mu}(\qq) = \sum_{s\alpha} \sqrt{M_s} \epsilon_{\mu s}^{\alpha *} 
(\qq)
u_{s}^{\alpha} (\qq)$ is a normal coordinate, and
$f_{\mathbf{k}n}$ is the Fermi-Dirac distribution function for state $\ket{{\bf 
k}n}$.
Considering the large value of the Fermi energy compared to the 
phonon frequencies, the temperature dependence of 
$\Gamma^{\mathrm{el-ph}}_{\mu}(\qq)$
is very weak and the sum in Eq.~\eqref{hfhm-elph} is usually restricted to the
states at the Fermi surface~\cite{PhysRevB.6.2577}:
\beq
\Gamma^{\mathrm{el-ph}}_{\mu}(\qq) = \frac{2\pi \hbar 
\omega_{\mu}(\qq)}{N} \sum_{\mathbf{k} m n}
                      |g^{\mu}_{\mathbf{k}n,\mathbf{k}+\qq m}|^2
                      \delta(\epsilon_{\mathbf{k}n}) 
\delta(\epsilon_{\mathbf{k}+\qq m}).
\label{hfhm-elph-res}
\eeq
In this work we assume that the electron-phonon linewidth is temperature
independent and it will be calculated making use of Eq.~\eqref{hfhm-elph-res}.

As it can be noted from Eq.~\eqref{hfhm-elph-res}, 
$\Gamma^{\mathrm{el-ph}}_{\mu}(\qq)$
is independent of the phonon frequency. However, it depends on the phonon 
polarization
vectors. If perturbation theory breaks down, the harmonic polarization vectors
might not be correct, yielding to a wrong linewidth. This problem is overcome
substituting in Eq.~\eqref{hfhm-elph-res} the harmonic polarization vectors
with the SSCHA polarization vectors, allowing the calculation of electron-phonon
linewidths in strongly anharmonic crystals.    

%
%
\begin{table}[b!]
\caption{Calculated contribution of the electron-phonon and phonon-phonon 
interaction to
the phonon linewidth at different $\qq$ points for PdH. 
The modes are ordered with increasing 
frequency (1-3 Pd-character acoustic modes, 4-6 H-character optical modes). 
The $\qq$ points are given in Cartesian coordinates in 
units of $2\pi/a$. Phonon linewidths are given in cm$^{-1}$ and represent the 
HWHM.
}
\begin{center}
\begin{tabular*}{0.75\linewidth}{c c c c c }
\hline
\hline
$\qq$ & $\mu$ & $\Gamma^{\mathrm{el-ph}}_{\mu}(\qq)$ & 
\multicolumn{2}{c}{$\Gamma^{\mathrm{ph-ph}}_{\mu}(\qq)$} \\ 
             &       &                                             & 0 K & 295~K 
\\
\hline
$[0.5,0.0,0.0]$  \\
             & 1 &    0.010 &    0.000 &   0.005 \\
             & 2 &    0.010 &    0.000 &   0.005 \\
             & 3 &    0.136 &    0.003 &   0.198 \\
             & 4 &    1.288 &    0.123 &  24.993 \\
             & 5 &    1.288 &    0.123 &  24.993 \\
             & 6 &    2.265 &   17.284 &  44.423 \\
${\rm X}=[1.0,0.0,0.0]$  \\
             & 1 &    0.142 &    0.000 &   0.012 \\
             & 2 &    0.142 &    0.000 &   0.012 \\
             & 3 &    0.027 &    0.006 &   0.069 \\
             & 4 &    4.212 &    0.123 &  24.152 \\
             & 5 &    4.121 &    0.123 &  24.152 \\
             & 6 &    3.202 &    1.496 &  13.166 \\
${\rm L}=[0.5,0.5,0.5]$  \\
             & 1 &    0.021 &    0.000 &   0.298 \\
             & 2 &    0.021 &    0.000 &   0.298 \\
             & 3 &    0.025 &    0.029 &   0.518 \\
             & 4 &    2.498 &    8.303 &  30.590 \\
             & 5 &    2.498 &    8.303 &  30.590 \\
             & 6 &    5.525 &    3.913 &   8.336 \\
\hline
\hline
\end{tabular*}
\end{center}
\label{table-gamma}
\end{table}

\begin{figure*}[t]
  \subfloat[PdH, 80~K]{\label{fig2a}
  \includegraphics[width=0.32\textwidth]{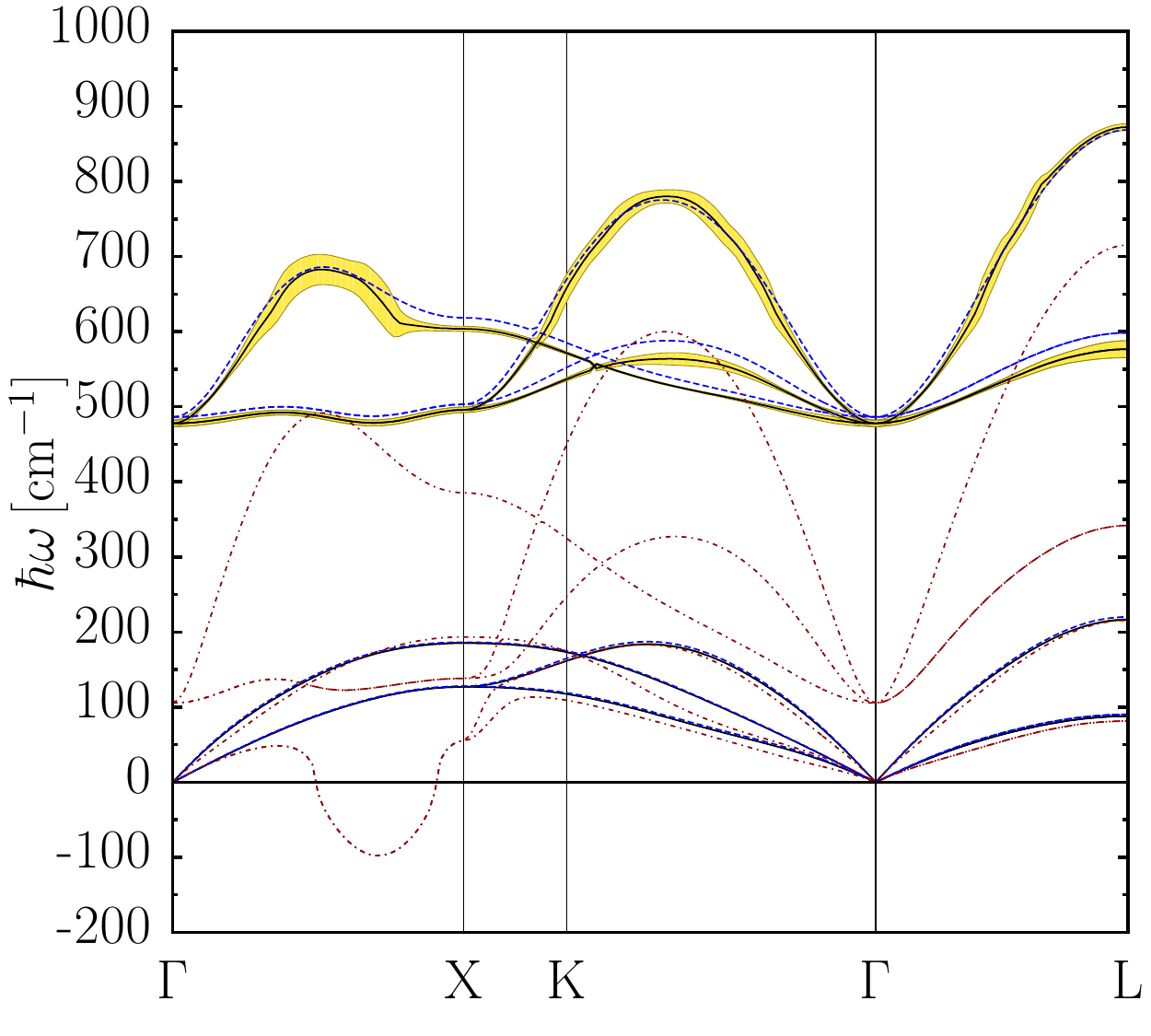}
  }
  \subfloat[PdD, 80~K]{\label{fig2b}
  \includegraphics[width=0.32\textwidth]{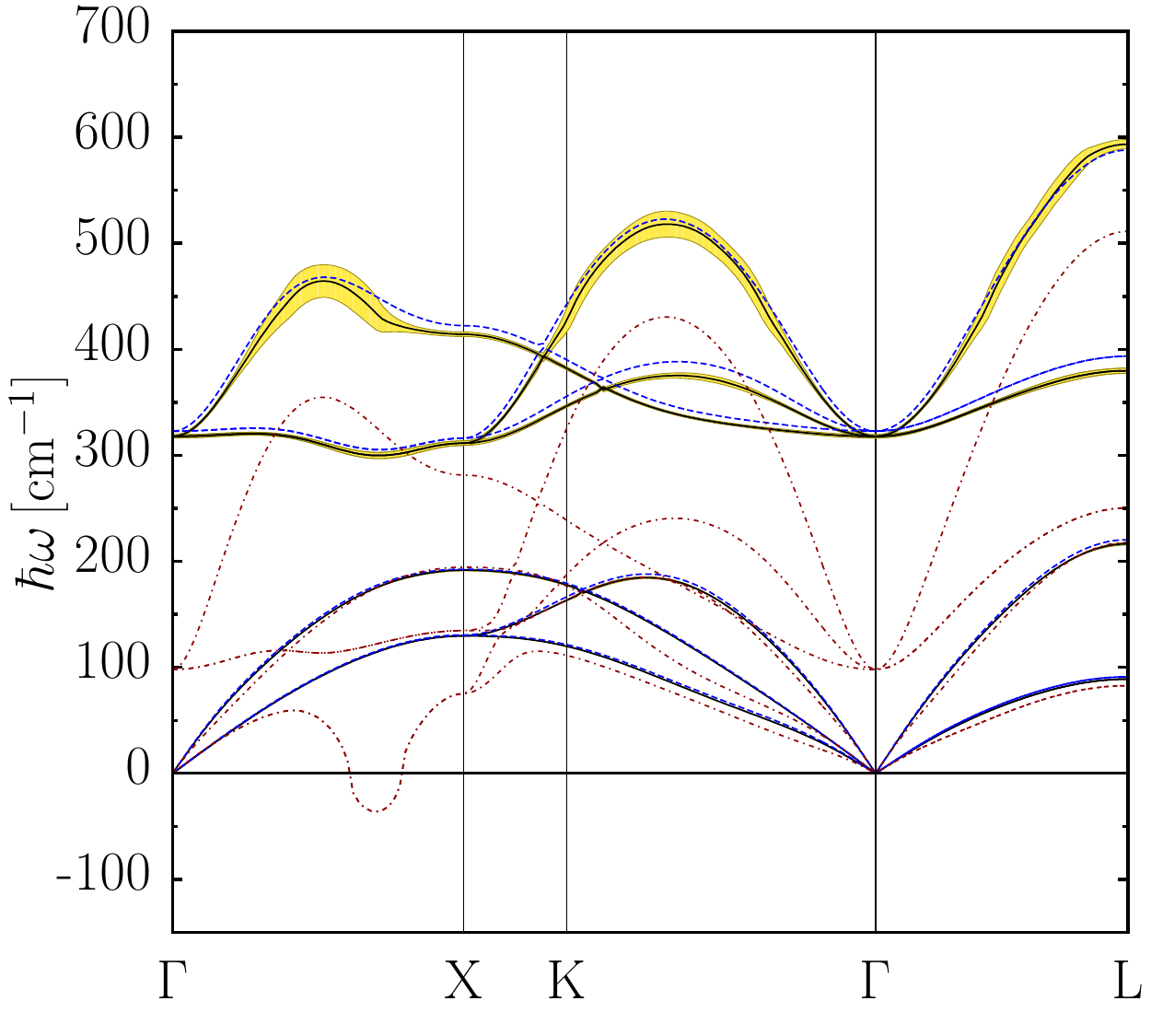}
  }
  \subfloat[PdT, 80~K]{\label{fig2c}
  \includegraphics[width=0.32\textwidth]{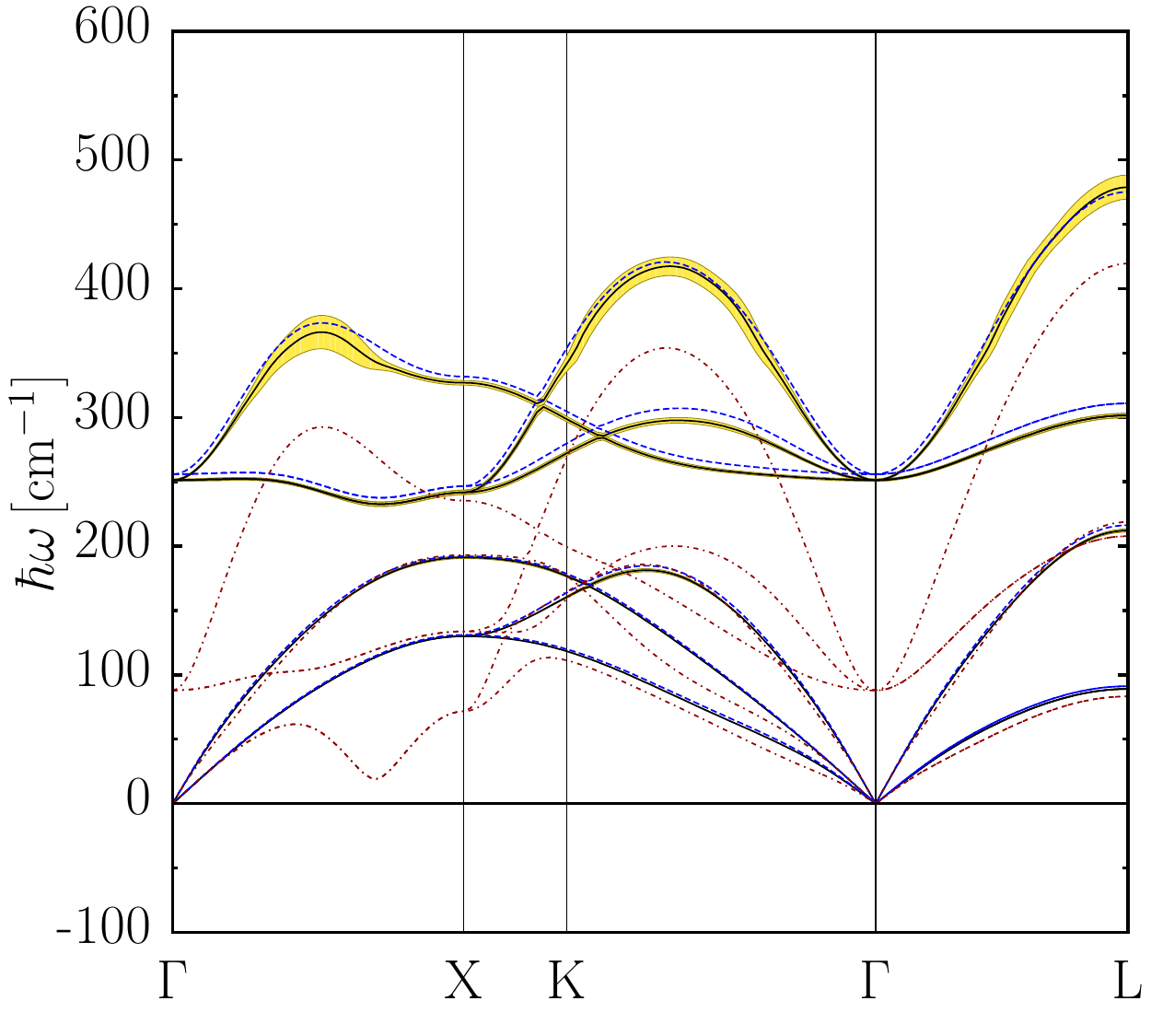}
  }
  \\
  \subfloat[PdH, 295~K]{\label{fig2d}
  \includegraphics[width=0.32\textwidth]{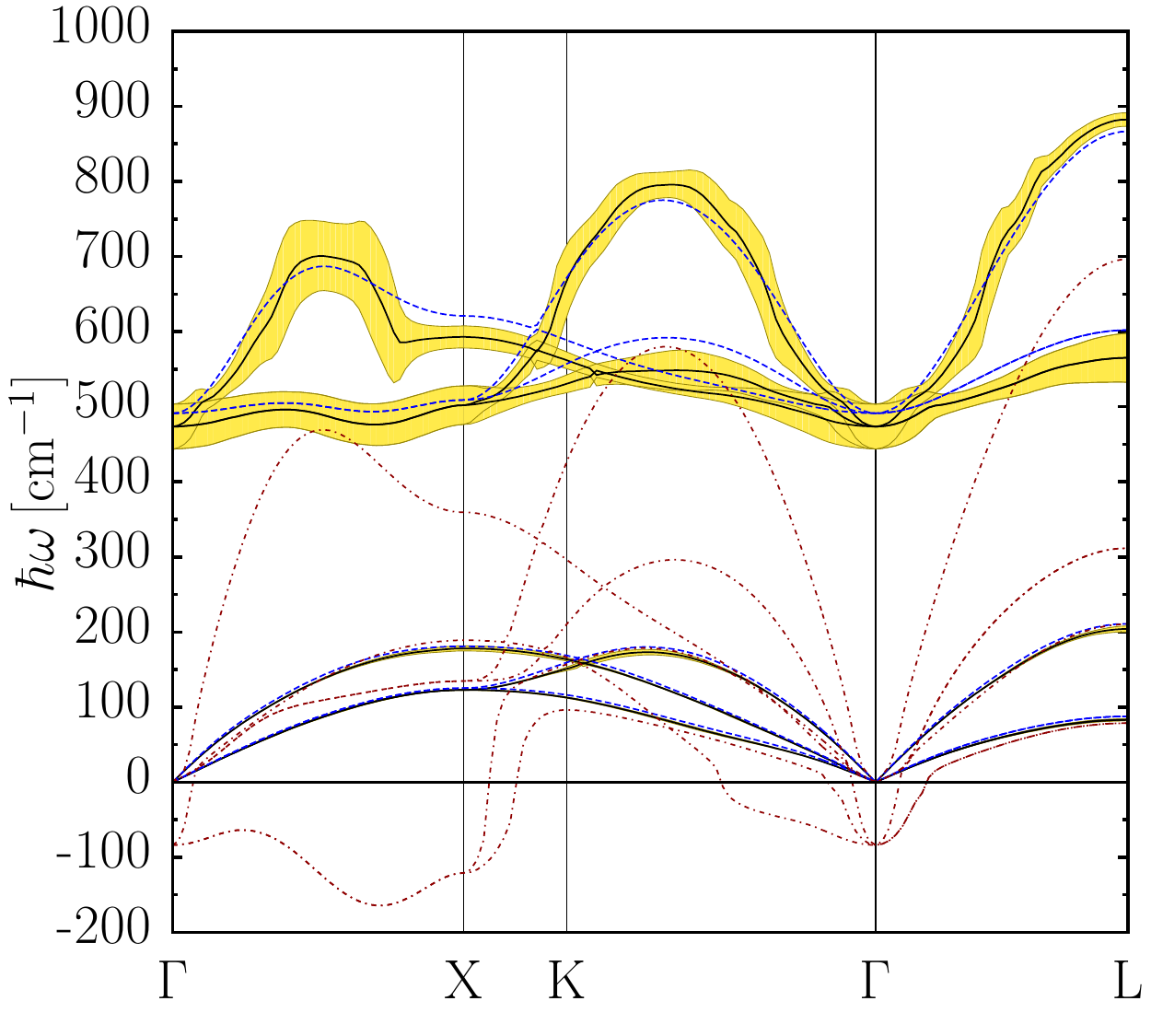}
  }
  \subfloat[PdD, 295~K]{\label{fig2e}
  \includegraphics[width=0.32\textwidth]{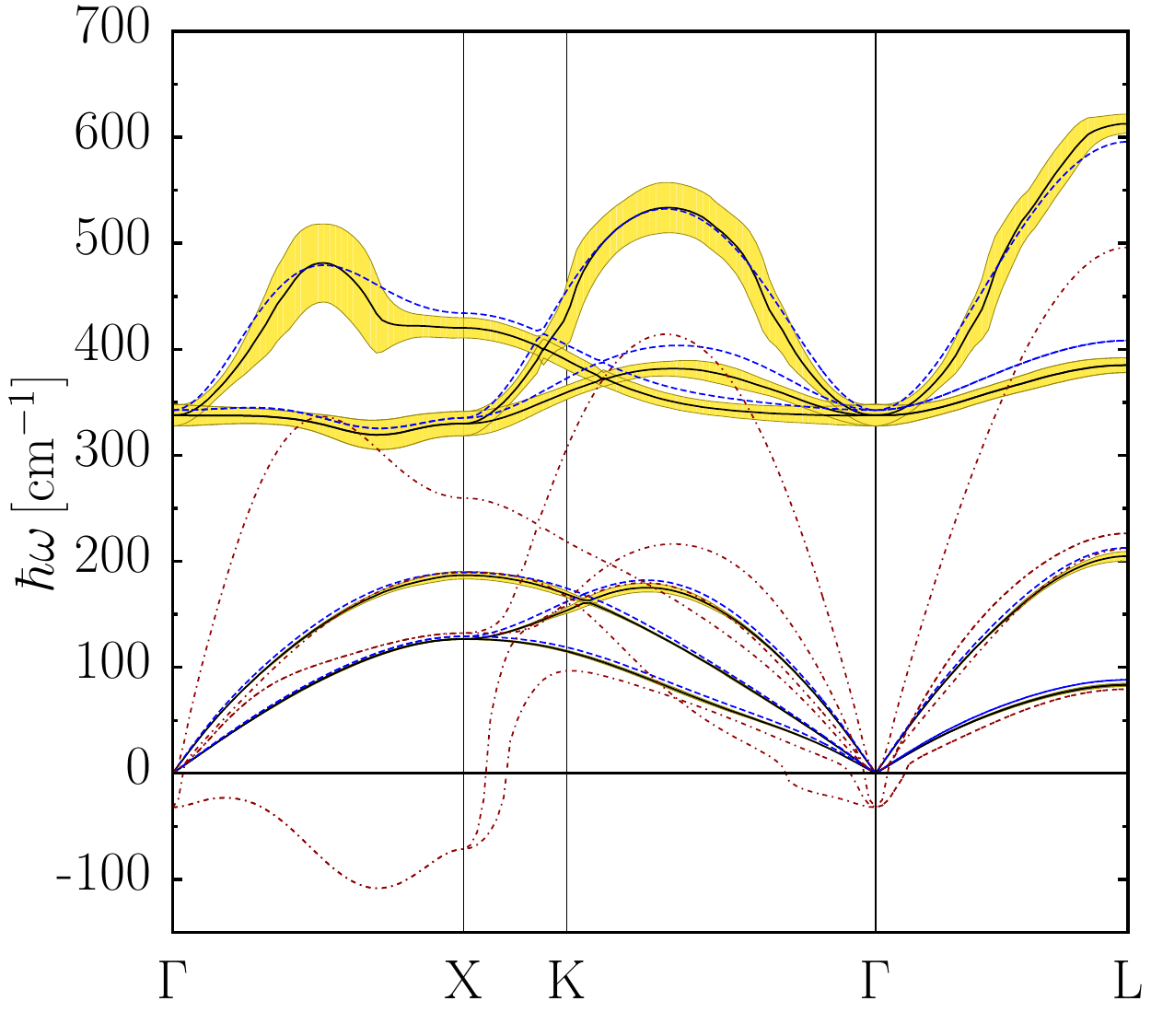}
  }
  \subfloat[PdT, 295~K]{\label{fig2f}
  \includegraphics[width=0.32\textwidth]{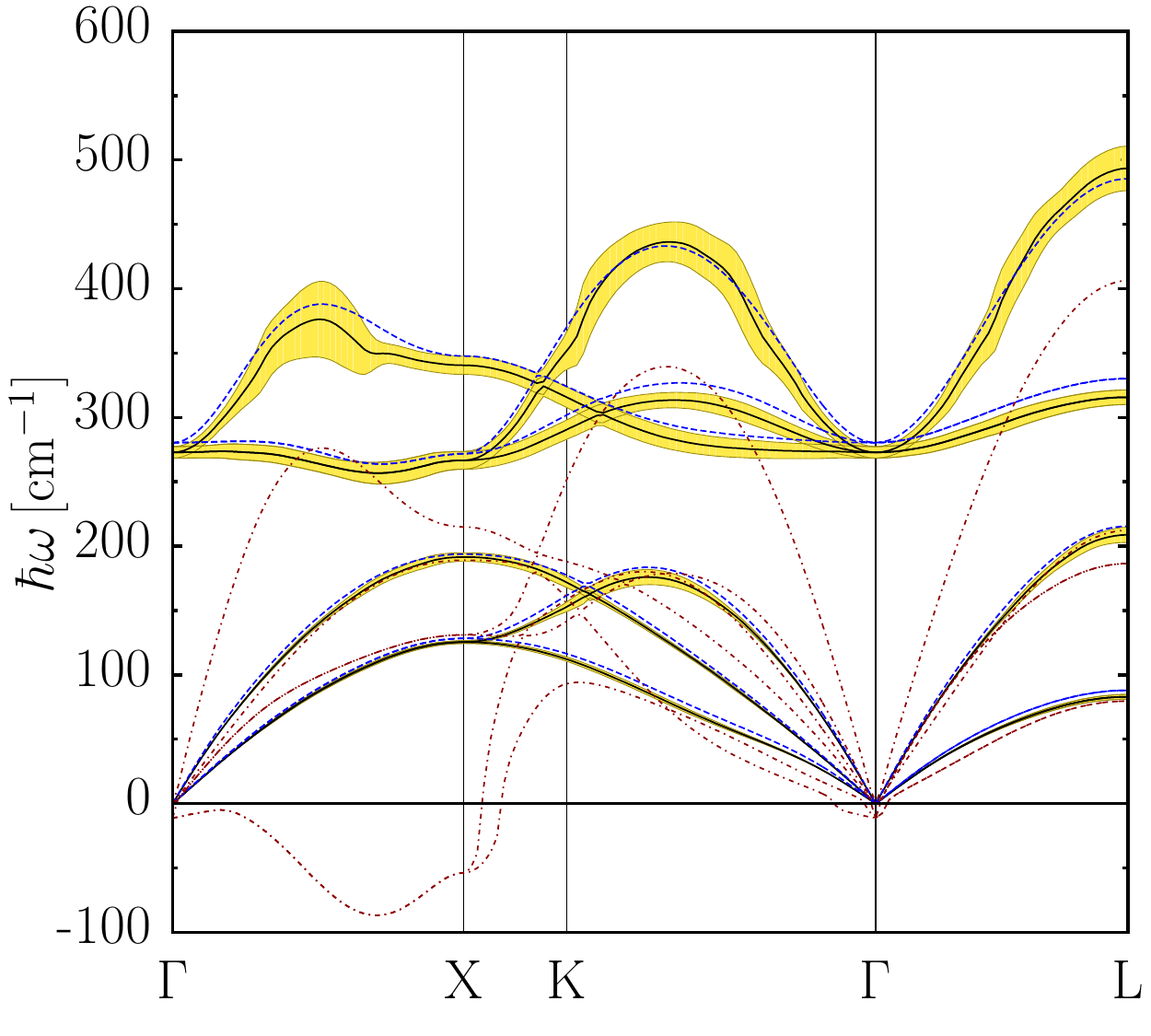}
  }
\caption{(Color online) Phonon spectra of palladium hydrides at 80 and 295~K. 
The red dashed-dotted line
is the harmonic phonon spectra including the thermal expansion, the blue dashed 
line
the SSCHA phonon spectra, and the black solid line represents the SSCHA spectra 
shifted with the
bubble contribution $\Delta_{\mu}(\qq)$ (see text). 
For the latter case the yellow width of each mode is the full width at
half maximum given by the anharmonic interaction, 
$2\Gamma^{\mathrm{ph-ph}}_{\mu}(\qq)$.}
\label{ph-spectra}
\end{figure*}

\begin{figure*}[t]
  \subfloat[PdH, 80~K]{\label{fig3a}
  \includegraphics[width=0.30\textwidth]{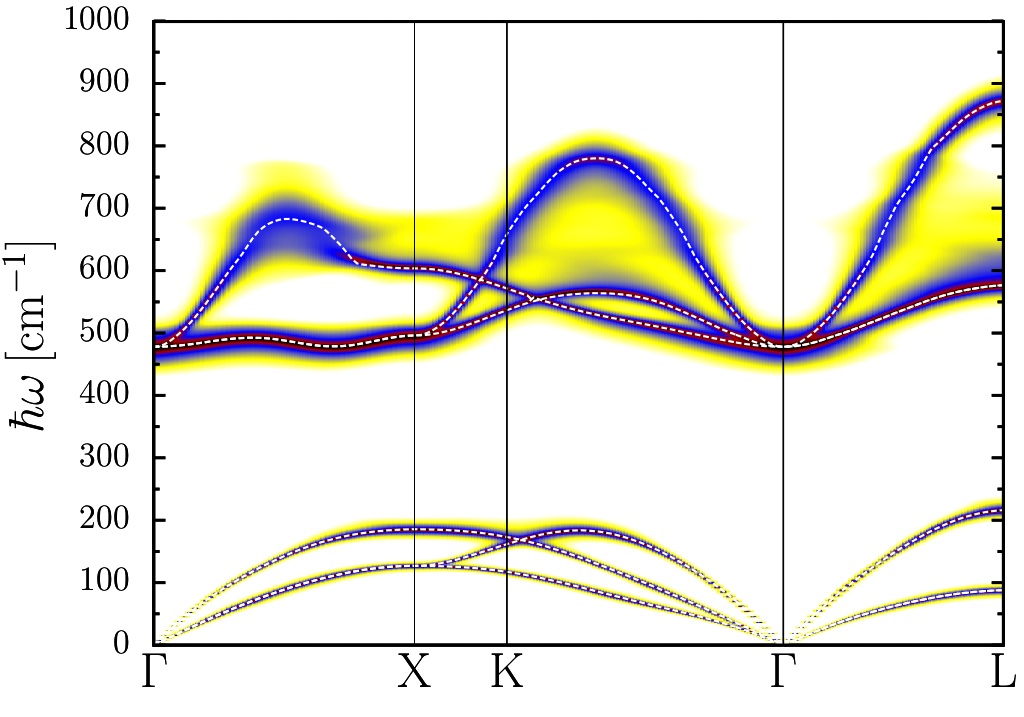}
  }
  \subfloat[PdD, 80~K]{\label{fig3b}
  \includegraphics[width=0.30\textwidth]{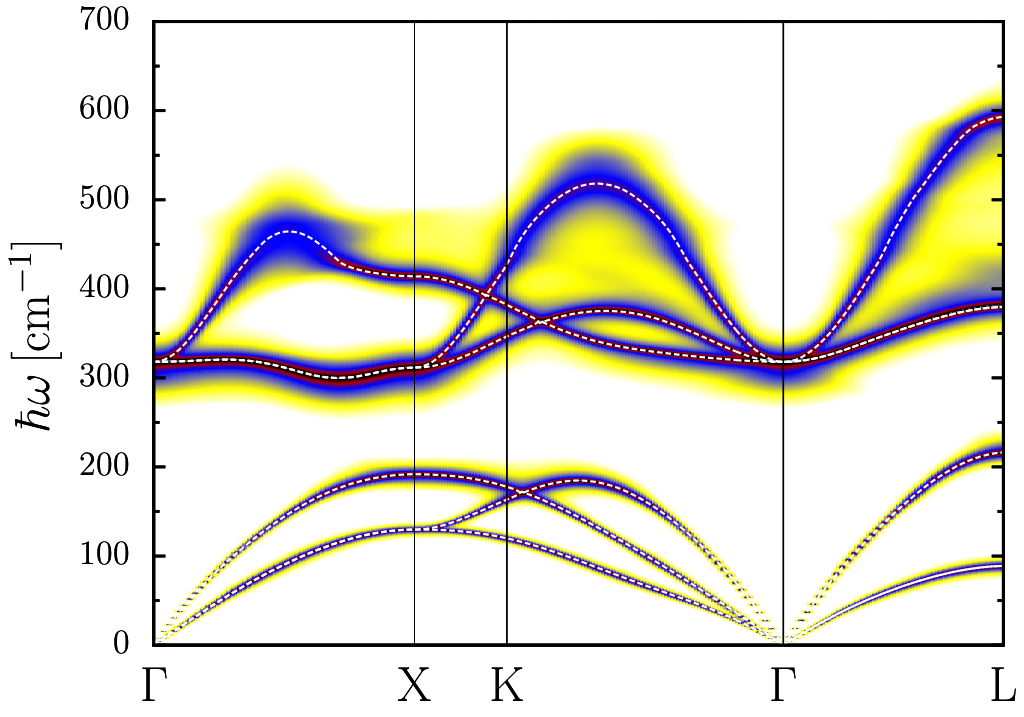}
  }
  \subfloat[PdT, 80~K]{\label{fig3c}
  \includegraphics[width=0.30\textwidth]{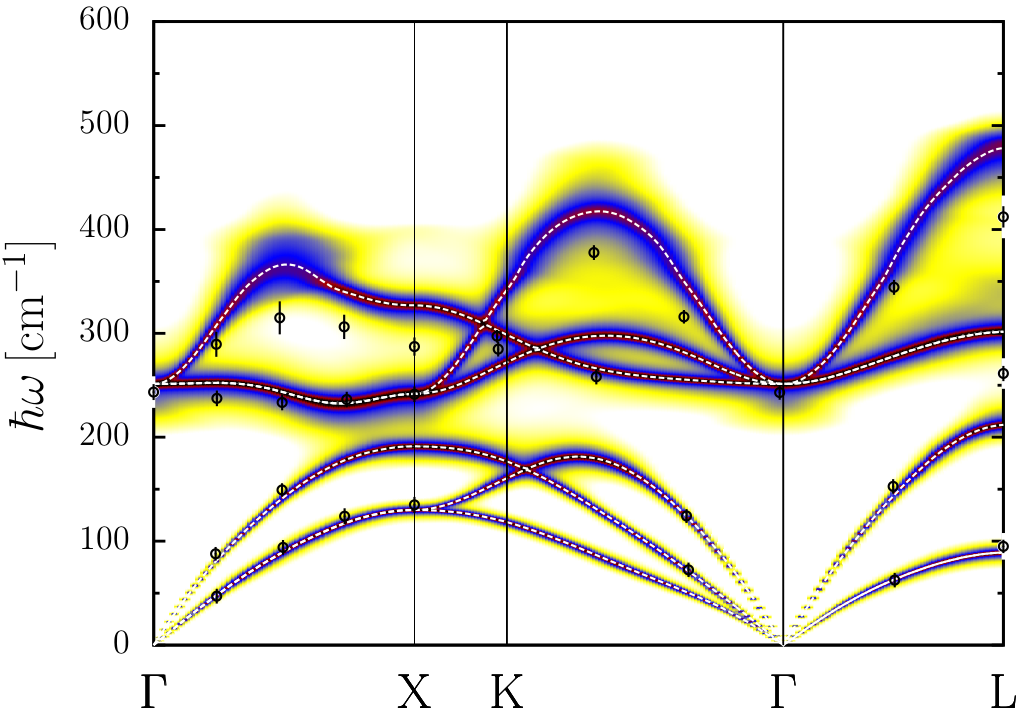}
  }
  \subfloat{
  \includegraphics[width=0.0676\textwidth]{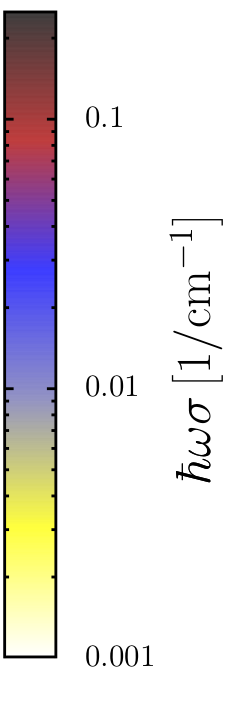}
  }
  \setcounter{subfigure}{3}
  \\
  \subfloat[PdH, 295~K]{\label{fig3d}
  \includegraphics[width=0.30\textwidth]{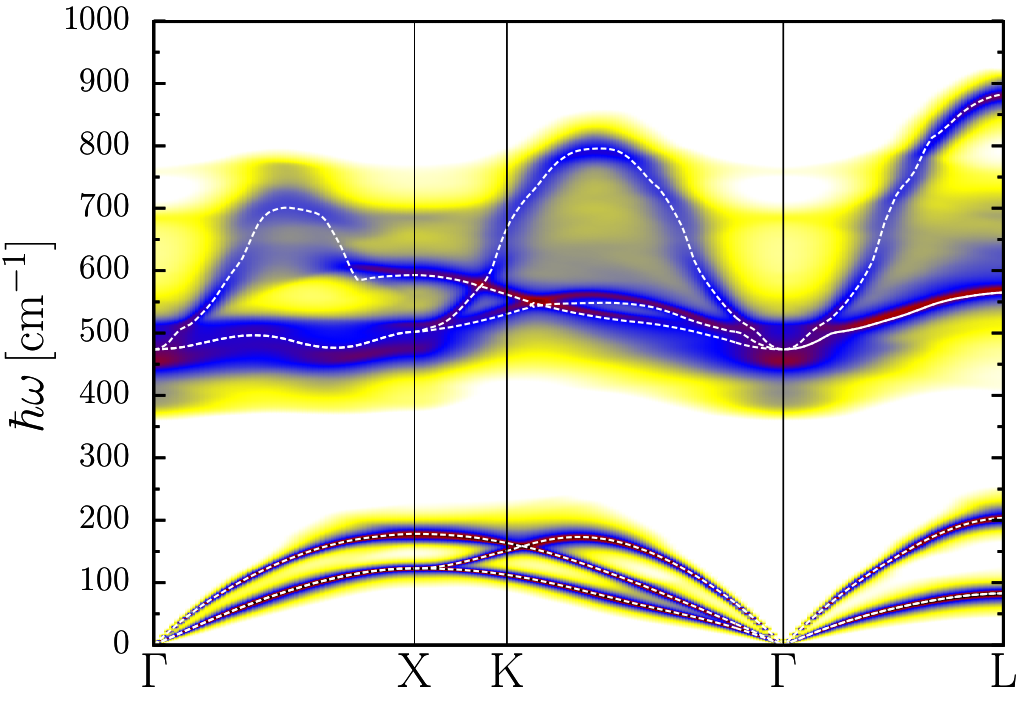}
  }
  \subfloat[PdD, 295~K]{\label{fig3e}
  \includegraphics[width=0.30\textwidth]{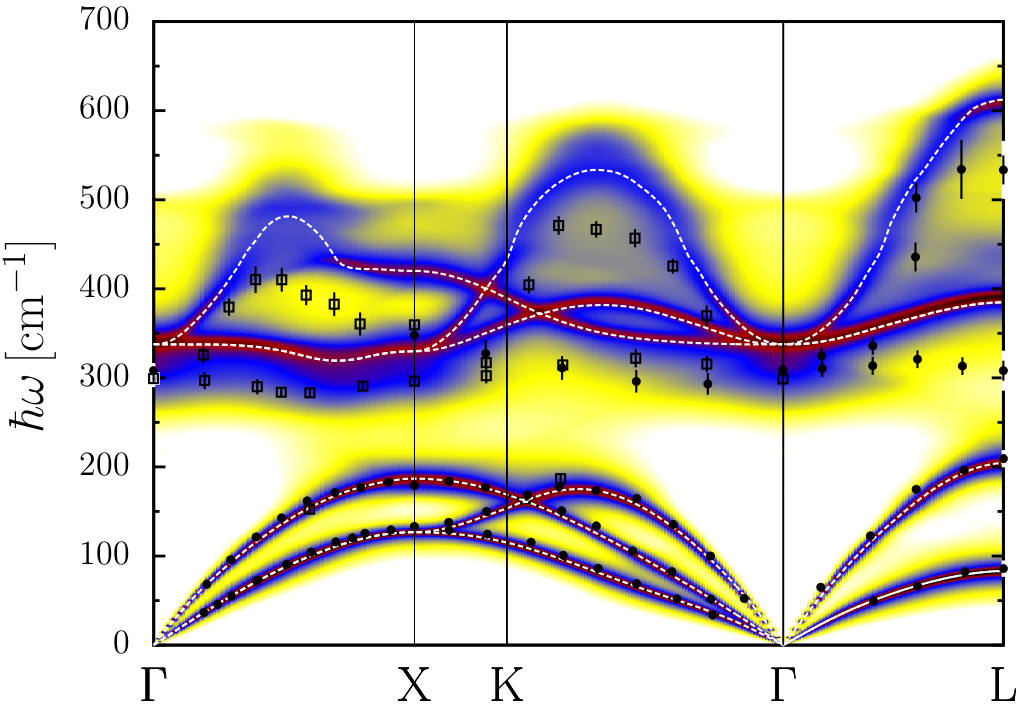}
  }
  \subfloat[PdT, 295~K]{\label{fig3f}
  \includegraphics[width=0.30\textwidth]{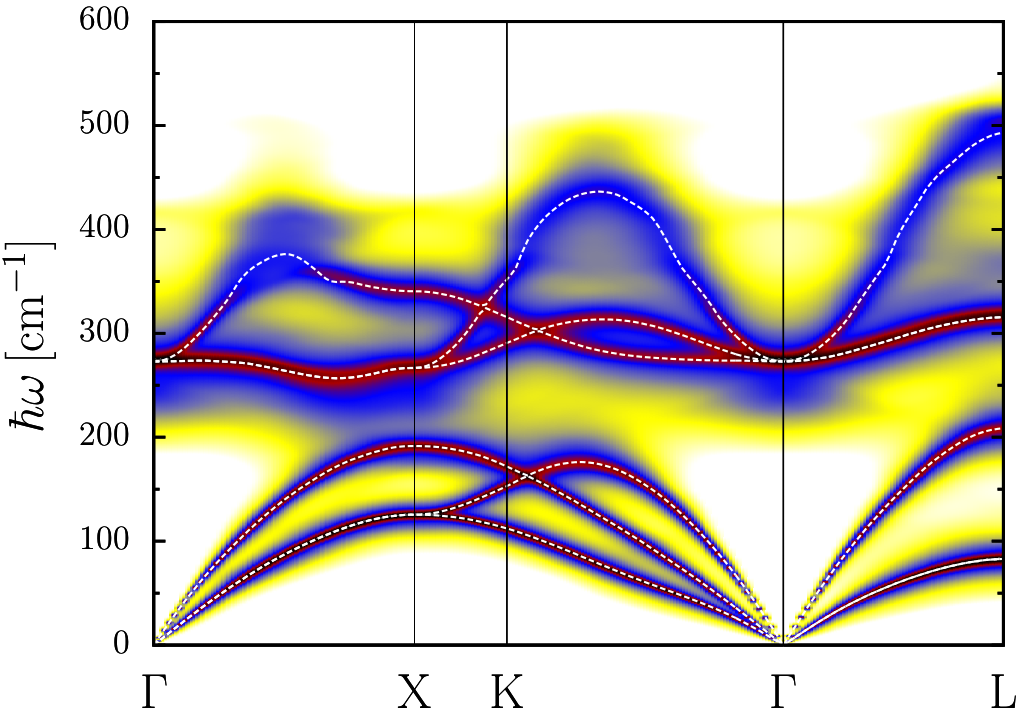}
  }
  \subfloat{
  \includegraphics[width=0.0676\textwidth]{thermo.png}
  }
\caption{
(Color online) Phonon spectra of palladium hydrides with full
inclusion of anharmonic effects. The white dashed lines represent the SSCHA phonon spectra
shifted by $\Delta_{\mu}(\qq)$. The color-shaded curves represent the
quantity
$\hbar \omega \sigma(\omega)$ (see Eq. \eqref{ins-sigma})  of palladium 
hydrides in logarithmic scale.
In (c) and (e) available experimental data for PdT$_{0.7}$ and PdD$_{0.63}$, 
respectively,
are plotted~\cite{PhysRevLett.33.1297,PhysRevLett.57.2955}.
}
\label{ph-spectral-function}
\end{figure*}

\begin{figure*}[t]
  \subfloat[PdH, 80~K, {\bf q}=$2\pi/a$ {[}1, 0, 0{]}]{\label{fig4a}
  \includegraphics[width=0.32\textwidth]{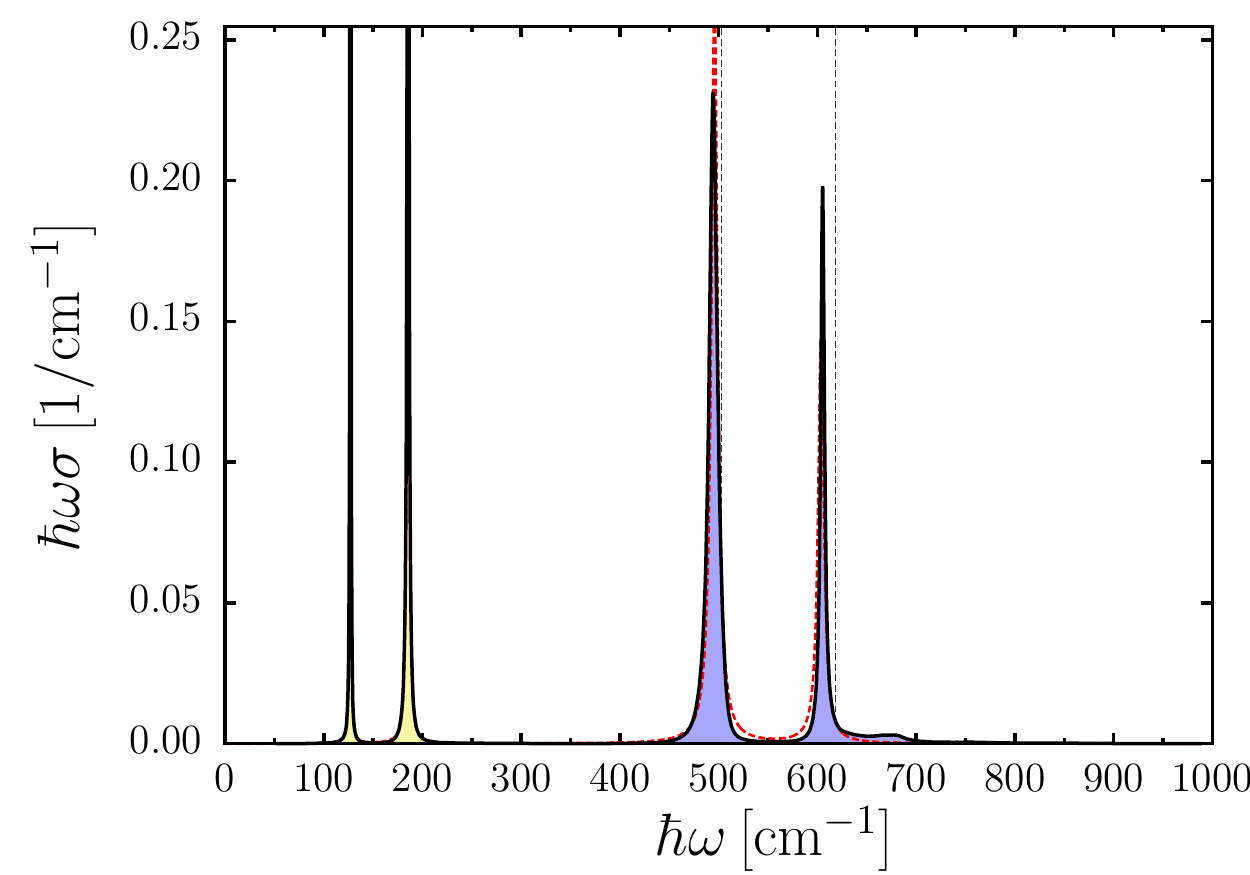}
  }
  \subfloat[PdD, 80~K, {\bf q}=$2\pi/a$ {[}1, 0, 0{]}]{\label{fig4b}
  \includegraphics[width=0.32\textwidth]{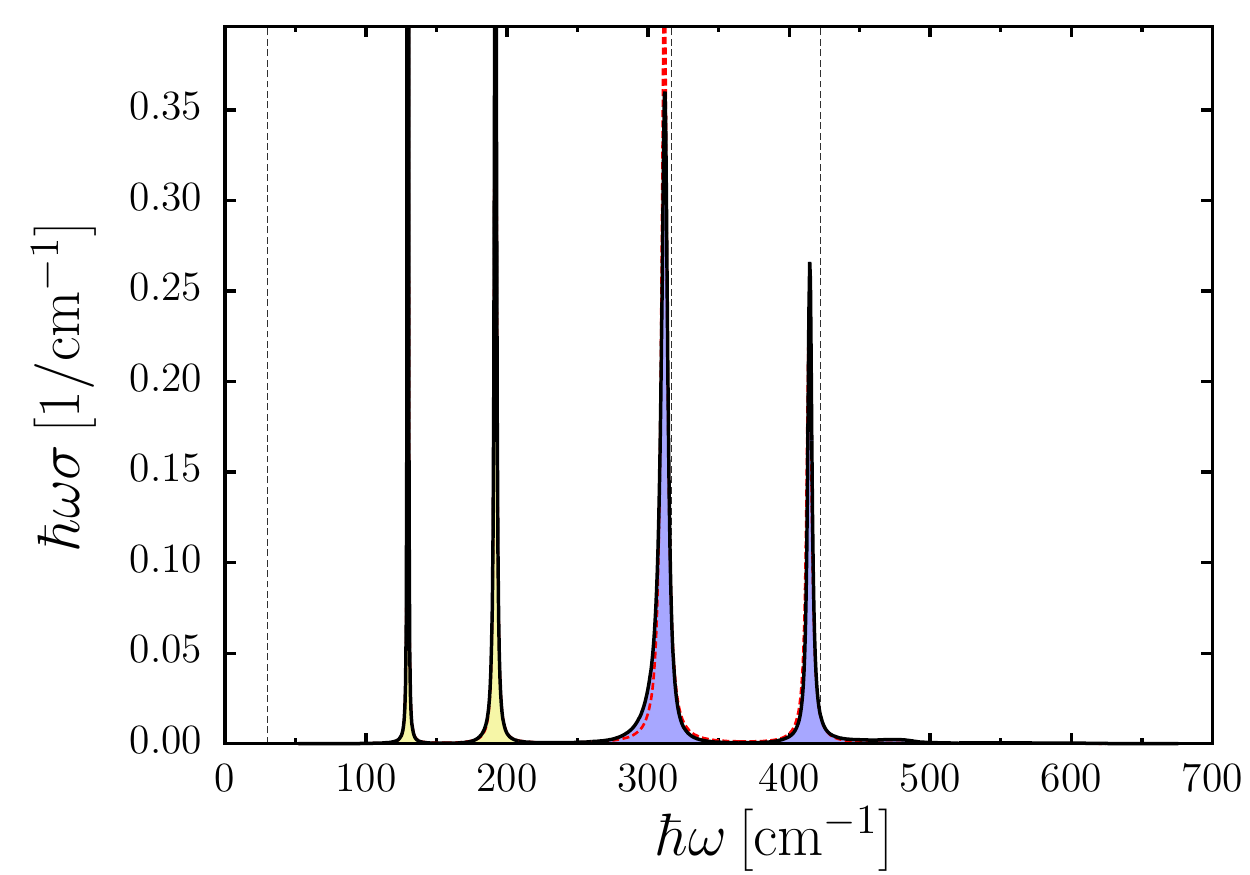}
  }
  \subfloat[PdT, 80~K, {\bf q}=$2\pi/a$ {[}1, 0, 0{]}]{\label{fig4c}
  \includegraphics[width=0.32\textwidth]{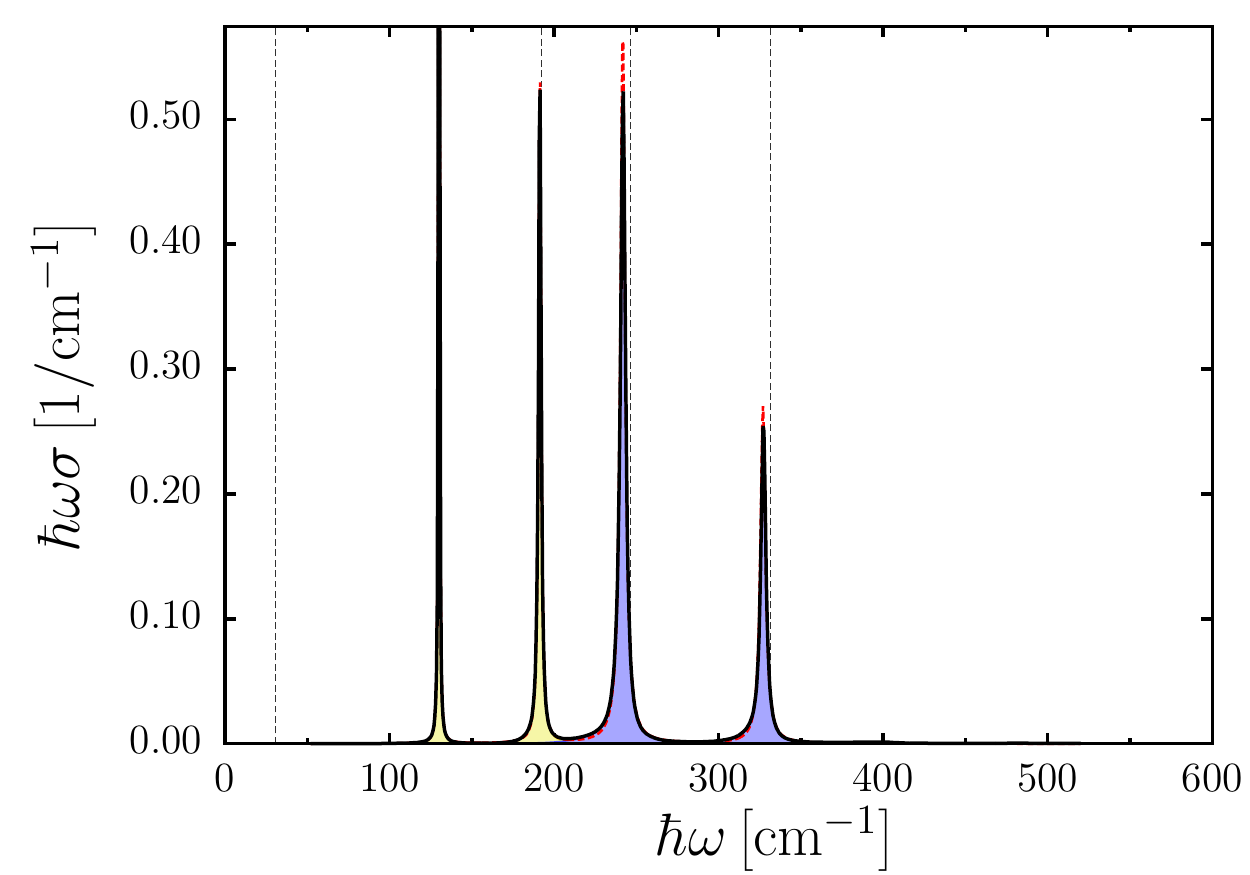}
  }
  \\
  \subfloat[PdH, 295~K, {\bf q}=$2\pi/a$ {[}1, 0, 0{]}]{\label{fig4d}
  \includegraphics[width=0.32\textwidth]{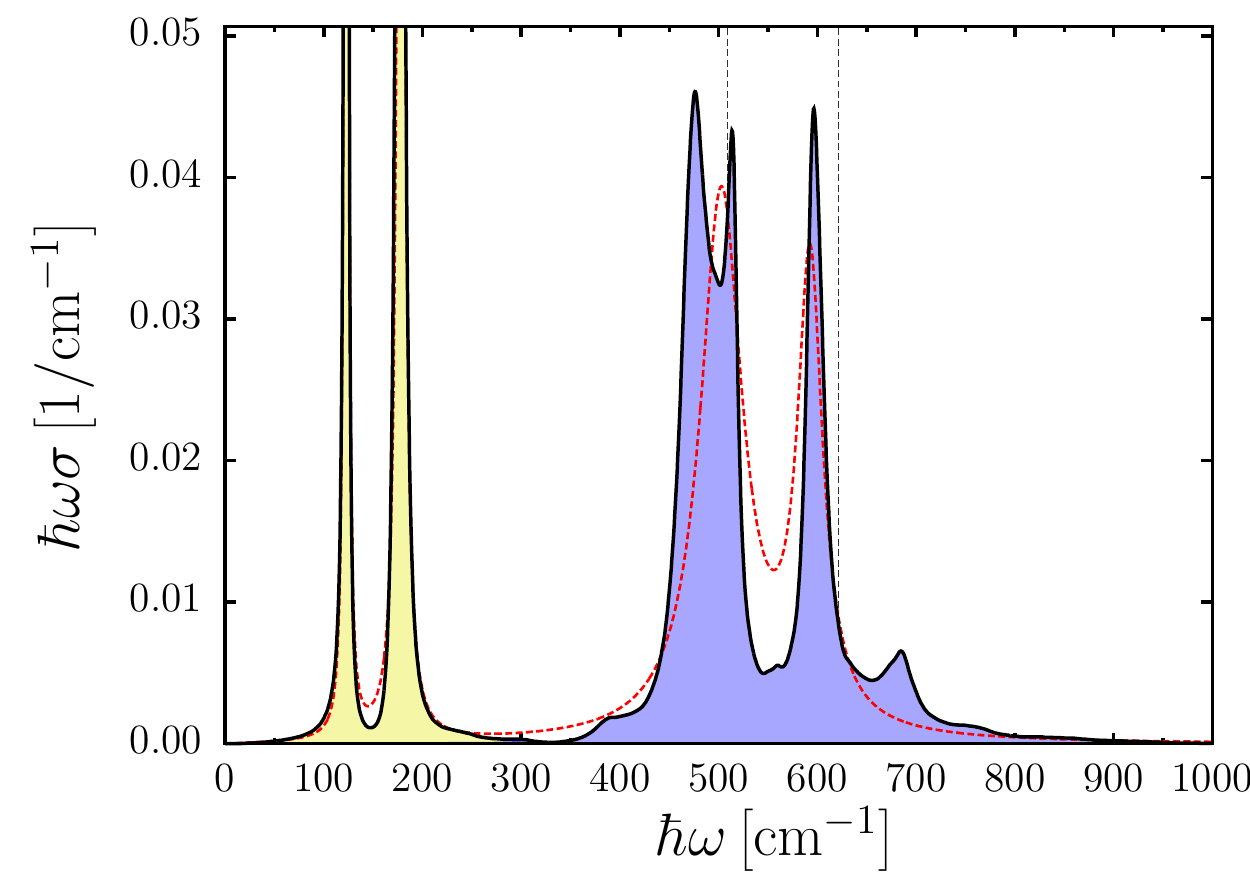}
  }
  \subfloat[PdD, 295~K, {\bf q}=$2\pi/a$ {[}1, 0, 0{]}]{\label{fig4e}
  \includegraphics[width=0.32\textwidth]{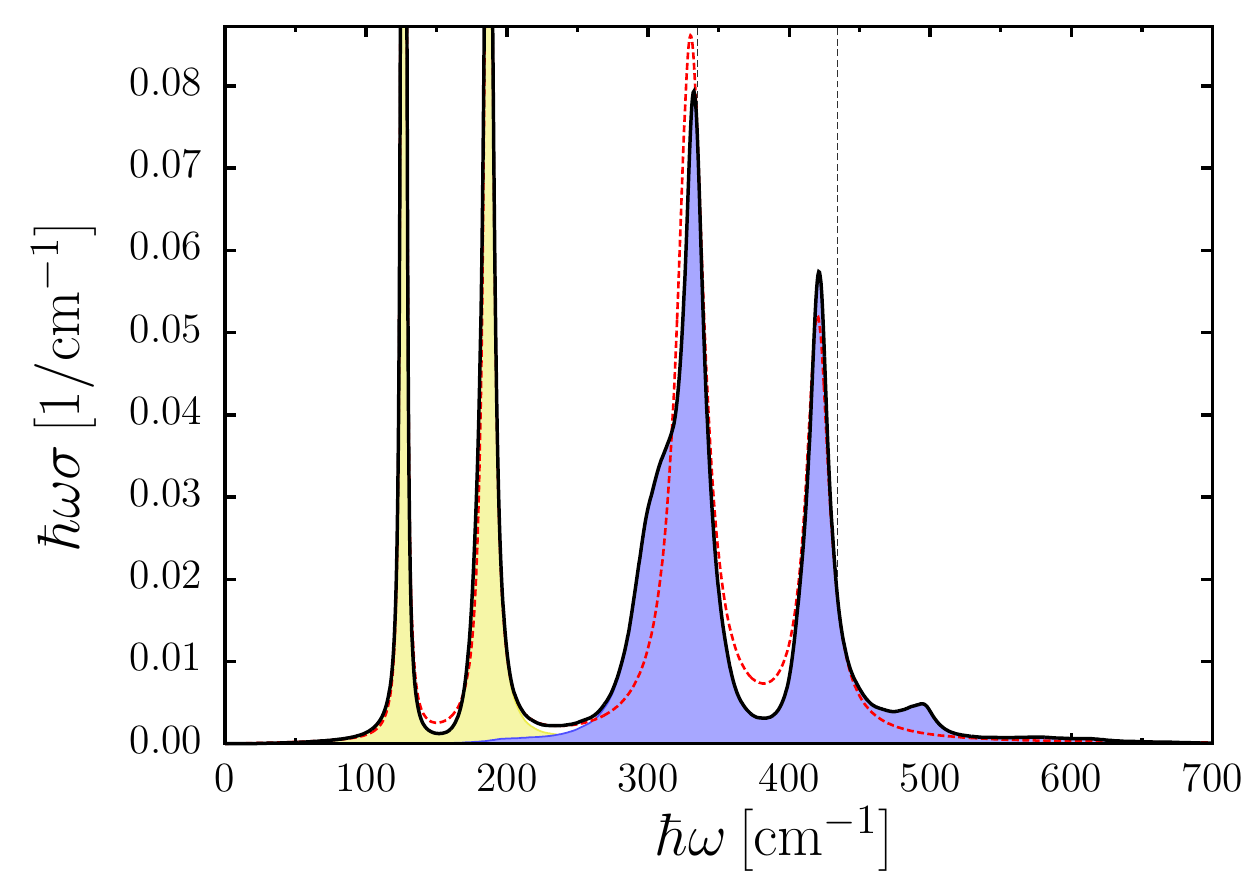}
  }
  \subfloat[PdT, 295~K, {\bf q}=$2\pi/a$ {[}1, 0, 0{]}]{\label{fig4f}
  \includegraphics[width=0.32\textwidth]{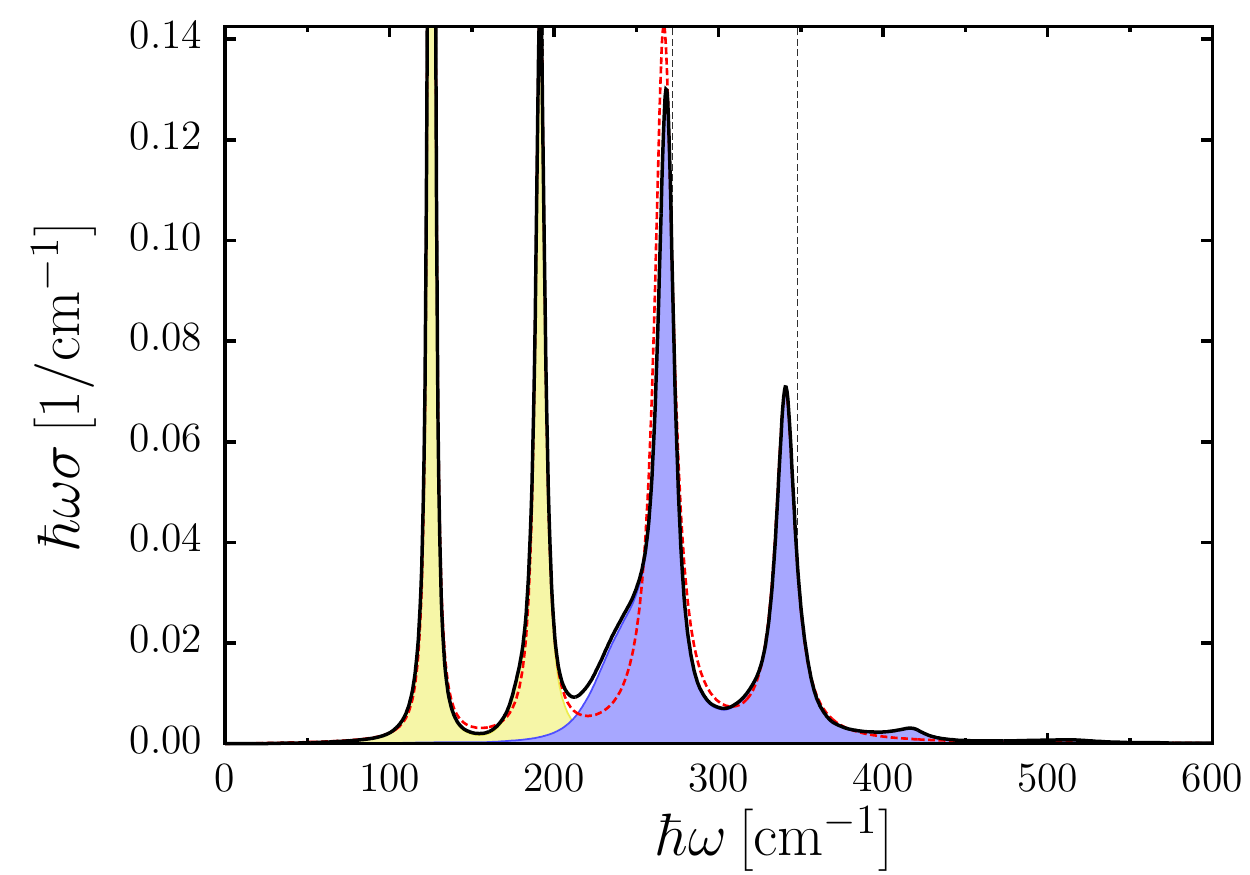}
  }
\caption{(Color online) $\hbar \omega \sigma(\qq, \omega)$ of palladium 
hydrides at 80 and 295~K
at the X point. The red dashed line represents the Lorentzian line-shape obtained substituting 
in Eq.~\eqref{ins-sigma}  
$\mathfrak{Im} \Pi^{\mathcal{H}B}_{\mu}(\qq,\omega) \sim - 
\Gamma^{\mathrm{ph-ph}}_{\mu}(\qq)$
and $\mathfrak{Re} \Pi^{\mathcal{H}B}_{\mu}(\qq,\omega) \sim 
\Delta_{\mu}(\qq)$.
The vertical black dashed lines denote the position of the SSCHA phonon 
frequencies. 
}
\label{ins-x}
\end{figure*}

\begin{figure*}[t]
  \subfloat[PdH, 80~K, {\bf q}=$2\pi/a$ {[}$\frac{1}{2}$, 0, 0{]}]{\label{fig5a}
  \includegraphics[width=0.32\textwidth]{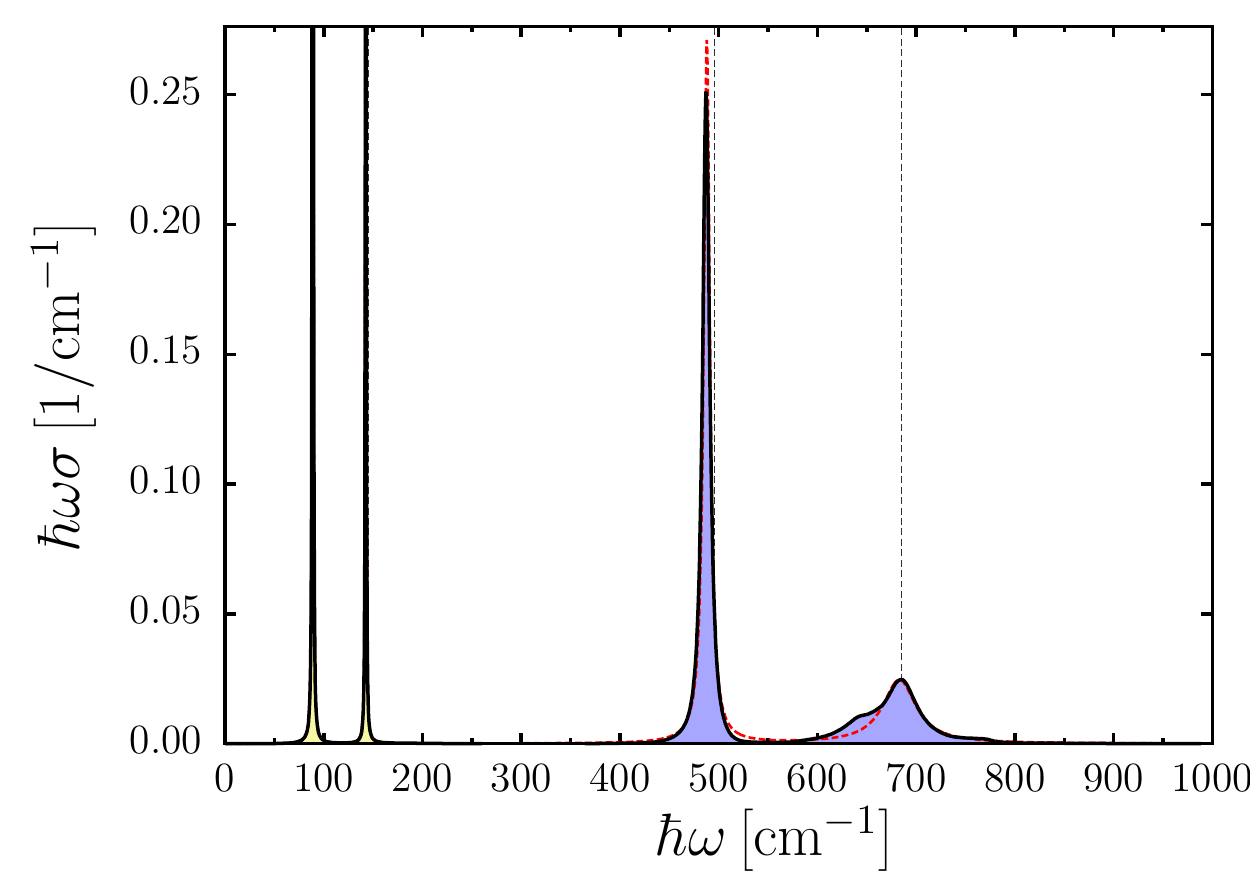}
  }
  \subfloat[PdD, 80~K, {\bf q}=$2\pi/a$ {[}$\frac{1}{2}$, 0, 0{]}]{\label{fig5b}
  \includegraphics[width=0.32\textwidth]{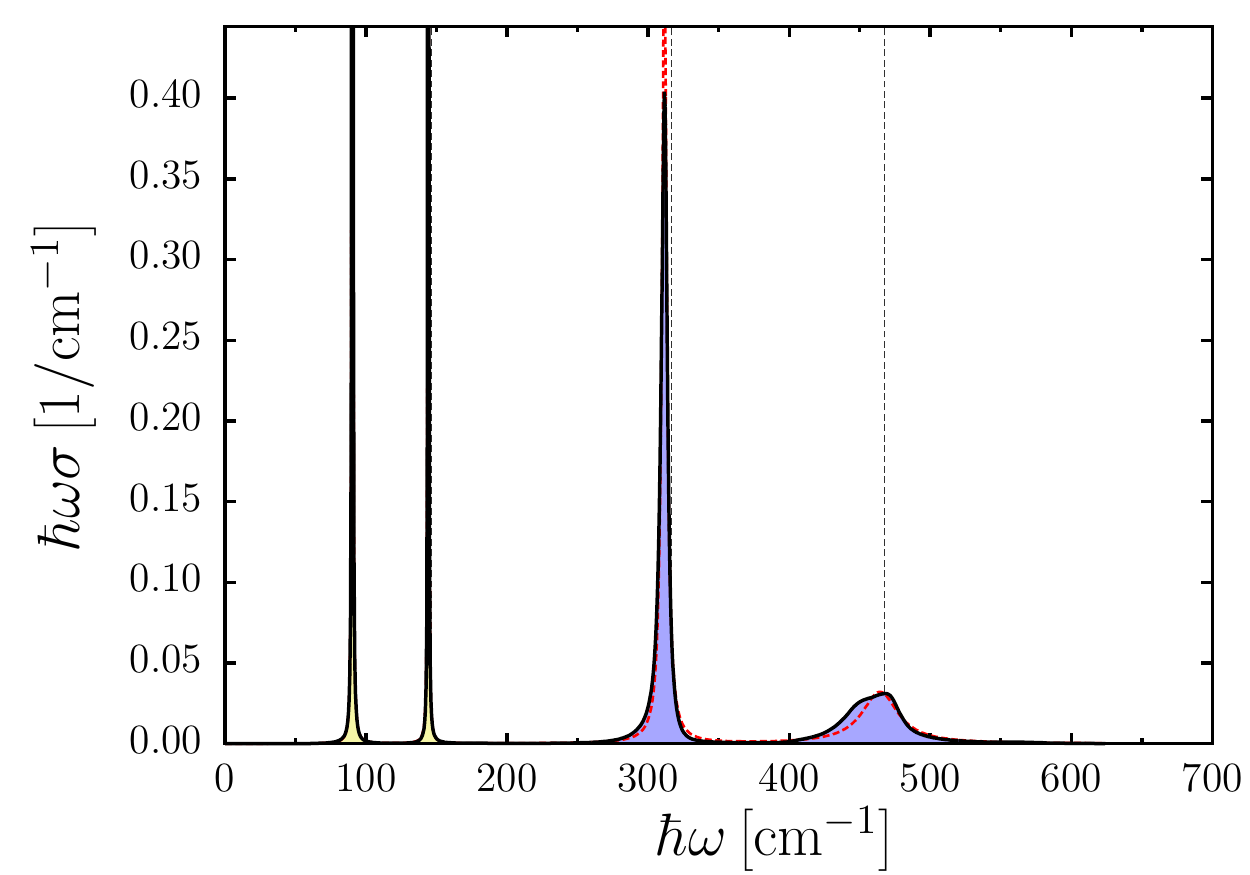}
  }
  \subfloat[PdT, 80~K, {\bf q}=$2\pi/a$ {[}$\frac{1}{2}$, 0, 0{]}]{\label{fig5c}
  \includegraphics[width=0.32\textwidth]{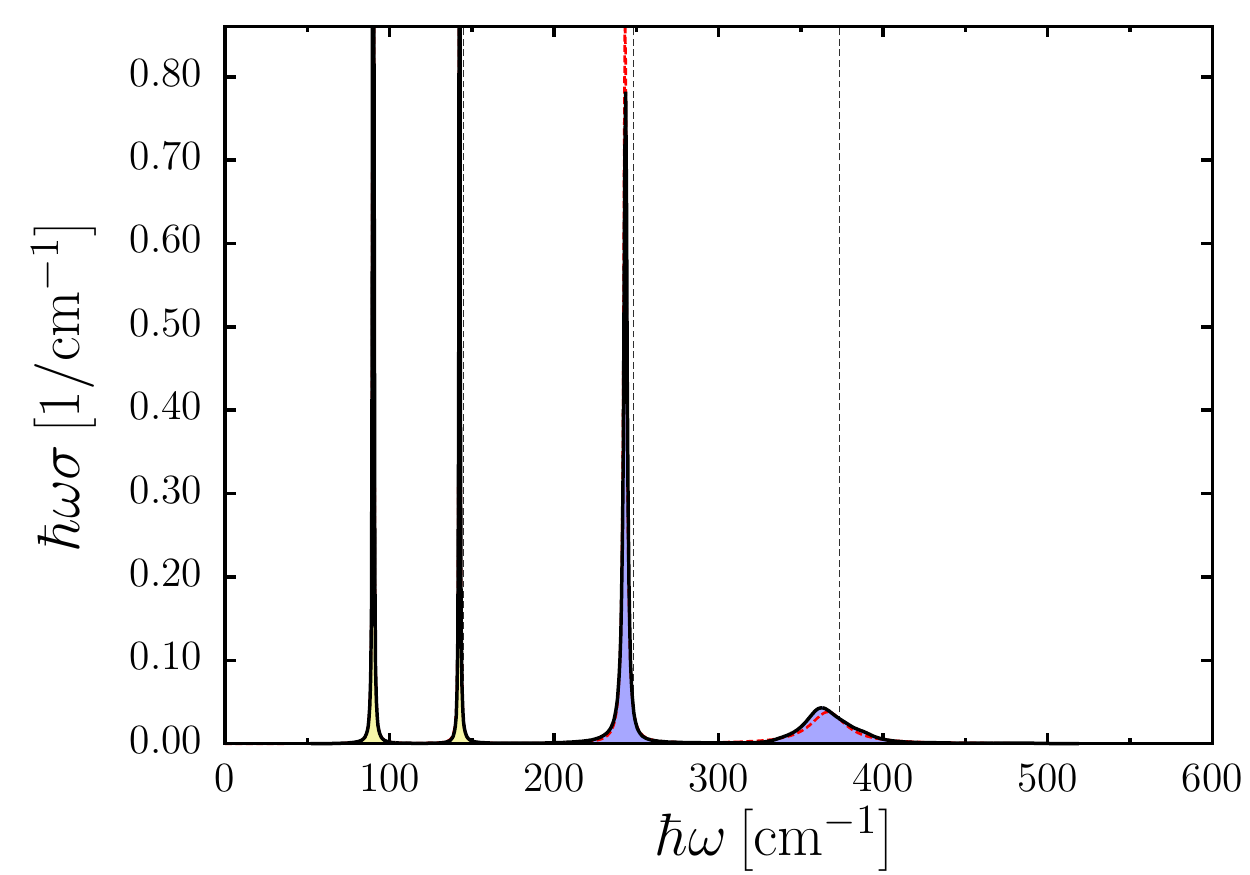}
  }
  \\
  \subfloat[PdH, 295~K, {\bf q}=$2\pi/a$ {[}$\frac{1}{2}$, 0, 0{]}]{\label{fig5d}
  \includegraphics[width=0.32\textwidth]{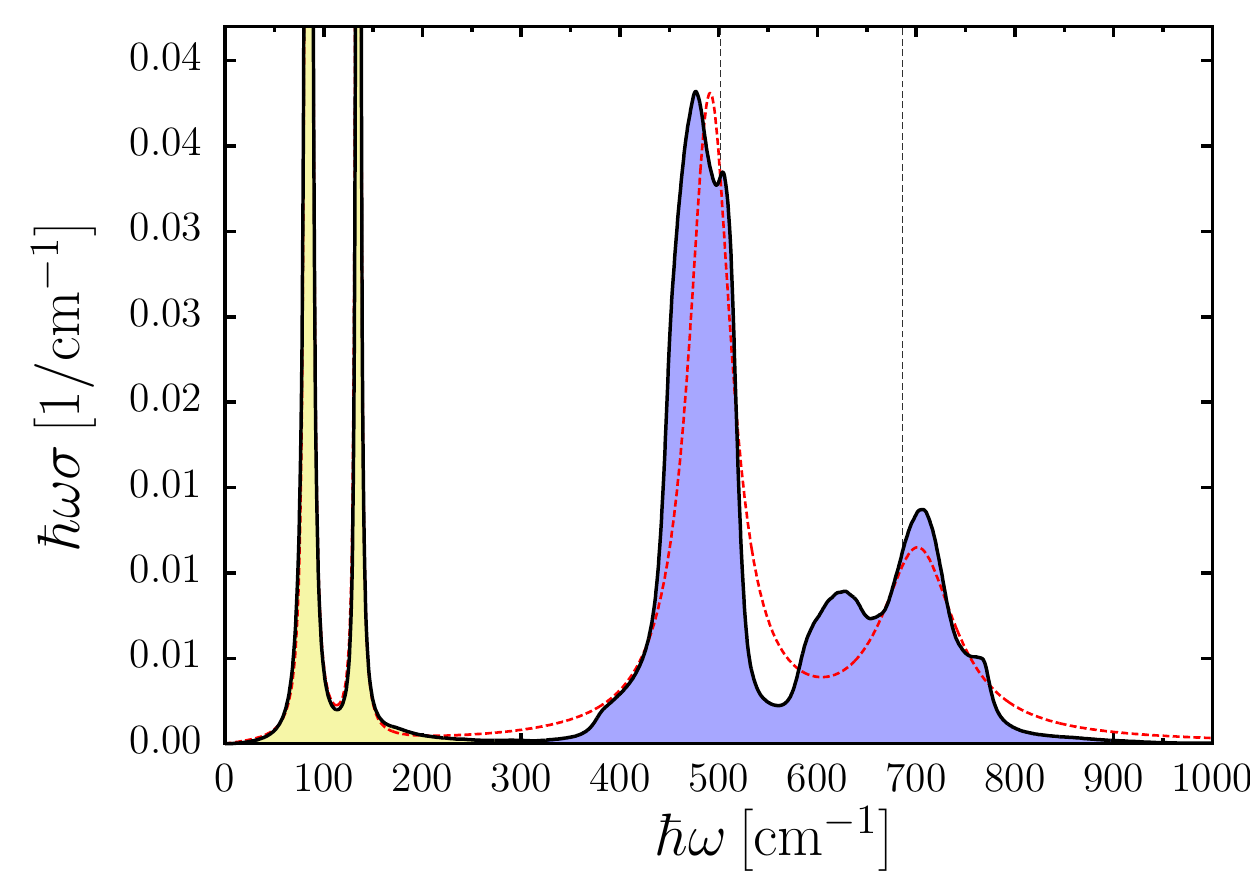}
  }
  \subfloat[PdD, 295~K, {\bf q}=$2\pi/a$ {[}$\frac{1}{2}$, 0, 0{]}]{\label{fig5e}
  \includegraphics[width=0.32\textwidth]{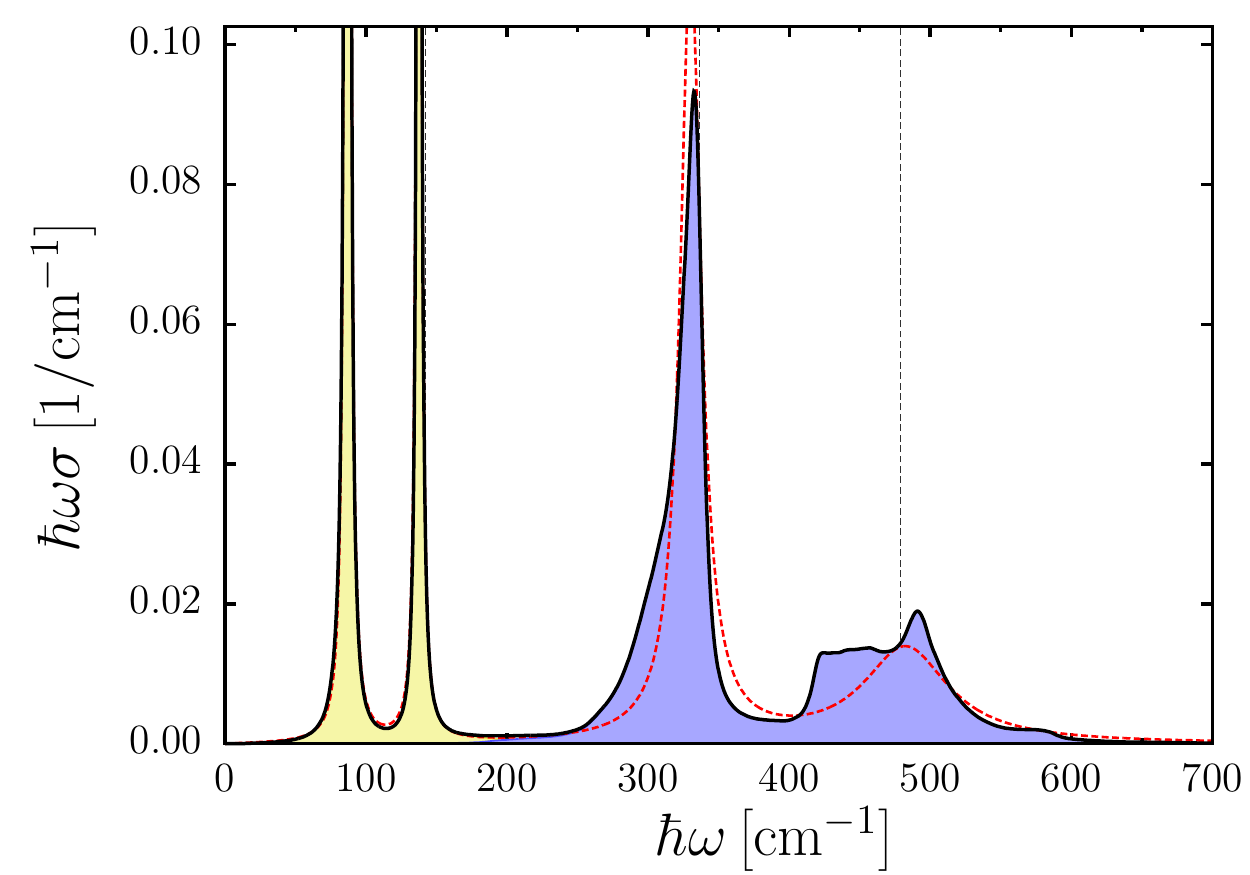}
  }
  \subfloat[PdT, 295~K, {\bf q}=$2\pi/a$ {[}$\frac{1}{2}$, 0, 0{]}]{\label{fig5f}
  \includegraphics[width=0.32\textwidth]{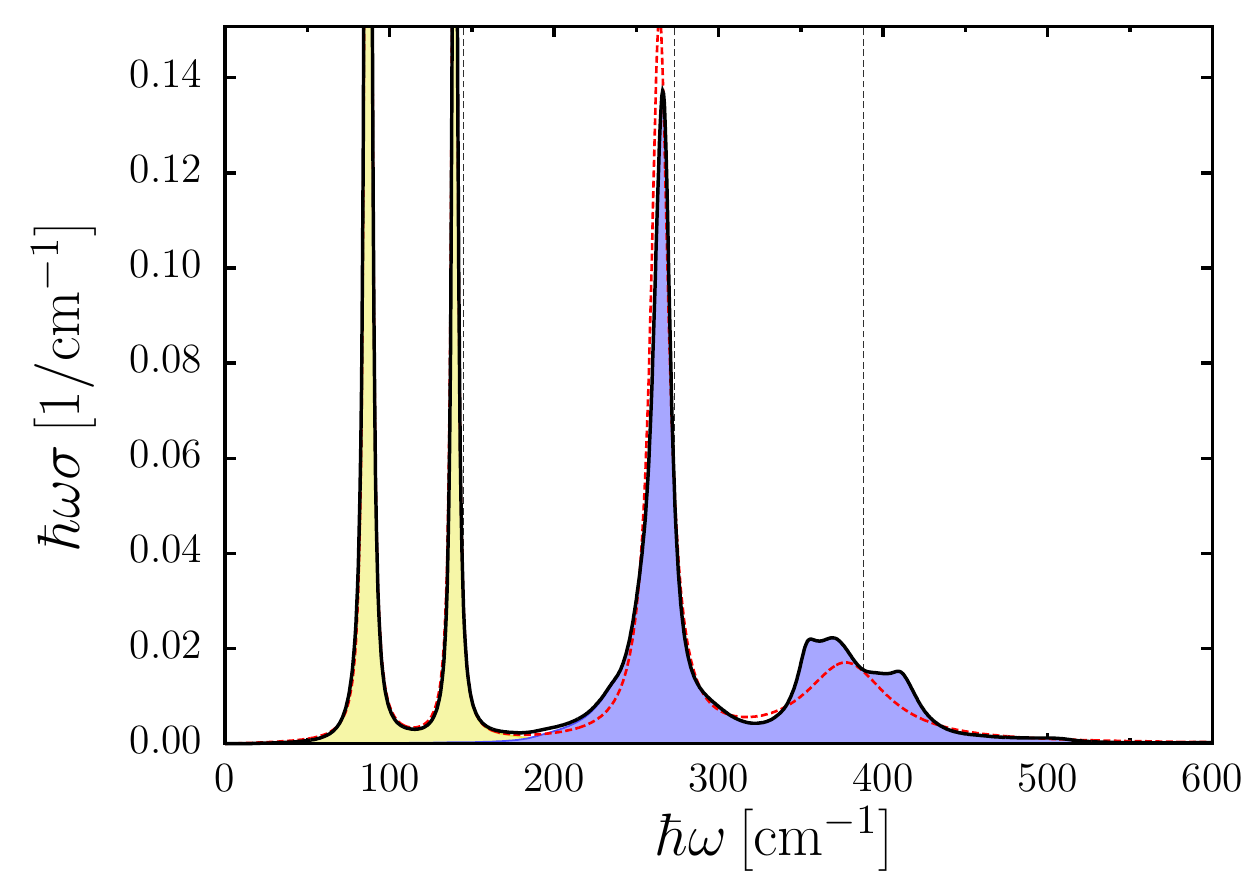}
  }
\caption{(Color online) $\hbar \omega \sigma(\qq, \omega)$ of palladium 
hydrides at 80 and 295~K
at $\qq=[1/2,0,0] 2\pi/a$. The red dashed lines represent 
the Lorentzian line-shape obtained substituting 
in Eq.~\eqref{ins-sigma}  
$\mathfrak{Im} \Pi^{\mathcal{H}B}_{\mu}(\qq,\omega) \sim - 
\Gamma^{\mathrm{ph-ph}}_{\mu}(\qq)$
and $\mathfrak{Re} \Pi^{\mathcal{H}B}_{\mu}(\qq,\omega) \sim 
\Delta_{\mu}(\qq)$.
The vertical black dashed lines denote the position of the SSCHA phonon 
frequencies. 
}
\label{ins-x2}
\end{figure*}
%


\section{Computational details}
\label{computational-details}

In this work we combine the calculation of the anharmonic
bubble diagram with the SSCHA as described in Sec.~\ref{scha-bubble} for 
palladium hydrides.
All calculations are performed within DFT making use the 
Perdew-Zunger local-density approximation~\cite{PhysRevB.23.5048}
and ultrasoft pseudopotentials. Harmonic phonon calculations are calculated
within DFPT as inplemented in {\sc
Quantum-ESPRESSO}~\cite{0953-8984-21-39-395502}. We have cut off the 
kinetic energy of the electron wavefunctions basis set at 50 Ry, 
and we have used a 24 $\times $ 24 $\times $ 24 Mokhorst-Pack 
mesh to sample the BZ.
The sum over $\protect \mathbf {k}$ in Eq.
\eqref{hfhm-elph-res} for the electron-phonon
linewidth requires a finer 72 $\times $ 72 $\times $ 72 grid.

The calculation of the third-order anharmonic coefficients 
$\phi_{s_1 s_2 s_3}^{\alpha_1 \alpha_2 \alpha_3} (\qq_1, \qq_2, 
\qq_3)$
is performed using the recent implementation of the 
 ``$2n+1$'' theorem~\cite{Paulatto13} for all the triplets of points 
with ${\bf q}_1$ and ${\bf q}_2$ in a $4 \times 4 \times 4$ regular mesh. 
The remaining $\qq_3$ is determined by crystal momentum conservation.
These coefficients are calculated
for the equilibrium volume calculated for PdH within 
the SSCHA~\cite{Errea13} at 0 K. We assume that
the $\phi_{s_1 s_2 s_3}^{\alpha_1 \alpha_2 \alpha_3} (\qq_1, 
\qq_2, \qq_3)$
coefficients do not change with temperature or isotope mass.
The coefficients were Fourier-transformed to three-body force constants and 
interpolated
on a finer grid of $30 \times 30 \times 30$ points
(Appendix C, Ref.~\onlinecite{Paulatto13}).
For the regularization $\delta$ of Eq. \eqref{f-fun} we have used 
5~cm$^{-1}$,
which corresponds to a Gaussian smearing of width of about 4~cm$^{-1}$
if the simplified formula of Eq.~(6) from 
Ref.~\onlinecite{Paulatto13} is used.

The SSCHA calculations have been performed as described in
Ref.~\onlinecite{Errea13}.
The temperature dependence of the SSCHA dynamical matrices
is estimated with the model potential
described in Ref.~\onlinecite{Errea13}. 
The thermal expansion is taken into account 
using the temperature 
dependent lattice parameters presented in
Ref.~\onlinecite{Errea14}.
However, we use the model potential exclusively to 
calculate the temperature shift of the dynamical
matrices as the zero-temperature dynamical matrices are
calculated fully \abinitio{} as in Ref.~\onlinecite{Errea13}. 
If the $D_{\mathrm{model}}(\qq,T)$ dynamical matrices
are those calculated with the model potential at temperature
$T$ and momentum $\qq$, 
the temperature-induced shift of the dynamical matrices is calculated
as $\Delta D(\qq,T) = D_{\mathrm{model}}(\qq,T) 
- D_{\mathrm{model}}(\qq,0)$. We add this difference 
to the $D(\qq,0)$ matrices calculated fully from first-principles
at 0 K, so that $D(\qq,T) = D(\qq,0) + \Delta D(\qq,T)$.


\section{Results and discussion}
\label{results}

The collapse of the harmonic approximation that was discussed in
Refs.~\onlinecite{Errea13} and~\onlinecite{Errea14} 
is evident in Fig.~\ref{ph-spectra}, where the harmonic phonon spectra
is plotted both at low and high temperature. The temperature dependence of the 
harmonic phonon spectra
comes solely from the thermal expansion. It is clear from the figure that at room 
temperature
the phonon instabilities become much more dramatic. 
The H-character optical modes become unstable even at the $\Gamma$
point, and strongly mix with the Pd-character acoustic modes
throughout the BZ. 
Considering that experimentally palladium hydrides exist even above 
room temperature in the rock-salt structure~\cite{Flanagan1991},
the instabilities predicted in the harmonic approximation 
are not real, and a non-perturbative scheme like the SSCHA is required
to study vibrational properties in palladium hydrides. 
   
In Fig.~\ref{ph-spectra} we present the SSCHA phonon spectra at
80 and 295~K, together with the SSCHA spectra shifted by 
$\Delta_{\mu}(\qq)$ calculated as described in Sec.~\ref{scha-bubble}.
The calculated $\Gamma^{\mathrm{ph-ph}}_{\mu}(\qq)$ linewidths are 
depicted as well.  
Within the SSCHA the energy of the H-character optical modes is strongly 
enhanced,
optical and acoustic modes disentangle, and the resulting
spectra is in rather good agreement with experimental 
results~\cite{Errea13,PhysRevLett.33.1297,PhysRevLett.57.2955}.
It is remarkable that the $\Delta_{\mu}(\qq)$ shift introduced
by the bubble diagram is very small compared to the SSCHA correction of 
the phonon frequency. Moreover, $\Delta_{\mu}(\qq)$
is small for all modes compared to the SSCHA renormalized 
$\Omega_{\mu}(\qq)$
frequencies. This is an \emph{a posteriori} confirmation
that the third-order term of the BO potential can be treated within
perturbation theory starting from the SSCHA solution
as discussed in Sec.~\ref{scha-bubble}.

The calculated phonon linewidth associated to the phonon-phonon
interaction is also small compared to the 
SSCHA phonon frequencies. This is expected as the real and imaginary
parts of the phonon self-energy are related by a Kramers-Kronig 
integral. As a consequence, despite the huge anharmonicity,
phonons in palladium hydrides 
are still quasiparticles that can be observed 
experimentally~\cite{PhysRevLett.33.1297,PhysRevLett.57.2955}.
This statement is still valid at high temperature, although
the linewidth associated to the phonon-phonon interaction
is strongly enhanced with temperature.

The values of the electron-phonon and phonon-phonon linewidths
are summarized in Table \ref{table-gamma}.
The $\Gamma^{\mathrm{ph-ph}}_{\mu}(\qq)$ linewidth is
larger than $\Gamma^{\mathrm{el-ph}}_{\mu}(\qq)$
for the optical H-character modes at most $\qq$ 
points even in the low-temperature regime. 
At 295~K the anharmonic linewidth is always larger than
the electron-phonon linewidth for these modes.
This is not the case for other superconducting metals,
where the linewidth coming
from the electron-phonon interaction is usually larger
than the anharmonic linewidth~\cite{Shukla03,PhysRevB.82.104504}.
The Pd-character acoustic modes have a considerably lower
anharmonic and electron-phonon linewidth,
remarking that they are very harmonic and barely
contribute to superconductivity~\cite{Errea13}.   

At low temperature, the phonon-phonon linewidth of the 
high-energy optical mode is larger than the linewidth of the low-energy
optical mode, especially halfway along $\Gamma$X. The reason
is that the high-energy optical phonons can decay into a low-energy 
optical mode and a high-energy acoustic mode while conserving
energy and crystal momentum. 
When temperature is increased
and acoustic modes start to be occupied, other optic modes
acquire a larger linewidth, but the one with largest
linewidth is still the highest-energy optical mode 
at $\qq=[1/2,0,0] 2\pi/a$, where $a$ is the lattice parameter. 

So far we have assumed that $\Gamma$ only depends on {\bf q} and on 
the phonon band, and that it is independent of the energy. 
This may however be false when the phonon linewidth becomes 
comparable with the energy separation of non-degenerate modes. 
However, this is not an issue for our technique, because 
having the SSCHA phonon frequencies and the
$\Pi^{\mathcal{H}B}_{\mu}(\qq,\omega)$ self-energy
we can simulate the phonon spectral function at an arbitrary 
{\bf q}-point, and without any assumption, according to the 
following expression:\cite{Cowley68}
\begin{widetext}
\beq
  \sigma(\qq, \omega) =  \sum_{\mu} \frac{ - 2 \hbar \Omega_{\mu} 
(\qq)
               \mathfrak{Im} \Pi^{\mathcal{H}B}_{\mu}(\qq,\omega) 
              }{
              [ \hbar^2 \omega^2 - \hbar^2 \Omega^2_{\mu} (\qq) - 2 \hbar 
\Omega_{\mu} (\qq)
              \mathfrak{Re} \Pi^{\mathcal{H}B}_{\mu}(\qq,\omega)
              ]^2
              + 4 \hbar^2 \omega^2_{\mu} (\qq)
               [\mathfrak{Im} \Pi^{\mathcal{H}B}_{\mu}(\qq,\omega)]^2
              }.
\label{ins-sigma}
\eeq
\end{widetext}
Substituting in Eq.~\eqref{ins-sigma}  
$\mathfrak{Im} \Pi^{\mathcal{H}B}_{\mu}(\qq,\omega) \sim - 
\Gamma^{\mathrm{ph-ph}}_{\mu}(\qq)$
and $\mathfrak{Re} \Pi^{\mathcal{H}B}_{\mu}(\qq,\omega) \sim 
\Delta_{\mu}(\qq)$
the spectral function $\sigma(\qq, \omega)$
reduces to a combination of Lorentzian functions.
This substitution is justified as long as the real part of the
self-energy remains constant in the energy range defined by the phonon linewidth.
Indeed, in most cases the measured inelastic neutron scattering (INS) spectra show 
Lorentzian line-shapes~\cite{Cowley68}, as the experimental 
phonon frequency is determined by the position
of the peak and the linewidth of the Lorentzian  
gives the experimental phonon linewidth.

In Fig.~\ref{ph-spectral-function} we plot the
spectral function $\sigma(\qq, \omega)$  
for PdH, PdD and PdT at 80 and 295~K, keeping the
full dependence on $\omega$ of the self-energy.
As it can be observed, while
the acoustic phonon peaks in $\sigma(\qq, \omega)$
fulfill with the simple Lorentzian picture,   
the highest energy modes already at 80~K show
shoulders in their peaks not expected \emph{a priori}.
The situation becomes more dramatic at high 
temperature, as $\sigma(\qq, \omega)$
shows very wide resonances with a non-Lorentzian
peak at most $\qq$ points for the optical modes,
sometimes even with satellite peaks. 
The reason for this is that the self-energy
is not constant in the range defined by the big 
linewidth of the optical modes. 
Having satellite peaks can be very misleading for experimentalists,
since they can be interpreted as structural phase transitions.
However, as in the case of palladium hydrides, secondary peaks
emerge due to strong anharmonicity. A similar effect has been recently
reported in PbTe~\cite{delaire614,PhysRevLett.112.175501}.

In Figs. \ref{ins-x} and \ref{ins-x2} we plot  $\hbar \omega \sigma(\qq, 
\omega)$
for the X point and $\qq=[0.5,0,0] 2\pi/a$, respectively.
We decided to plot $\sigma(\qq, \omega)$ multiplied by $\hbar \omega$ 
because the integral of this quantity over the entire energy domain 
gives the number of modes~\cite{0953-8984-9-8-005}.
The curve obtained substituting in Eq.~\eqref{ins-sigma}  
$\mathfrak{Im} \Pi^{\mathcal{H}B}_{\mu}(\qq,\omega) \sim - 
\Gamma^{\mathrm{ph-ph}}_{\mu}(\qq)$
and $\mathfrak{Re} \Pi^{\mathcal{H}B}_{\mu}(\qq,\omega) \sim 
\Delta_{\mu}(\qq)$
is also shown. It is clear from the figures that
the latter approximation yields good spectral peaks for 
the Pd-character acoustic modes, but not for the
H-character optical modes, specially at high temperature.
For instance, at the X point at 295~K for PdH 
we observe that the peak associated to the lowest-energy
optical mode is split into two distinct
peaks and the highest-energy optical mode
shows a satellite peak at high energy.
The departure from the Lorentzian line-shape is less acute
for the heavier isotopes. At $\qq=[0.5,0,0] 2\pi/a$
on the contrary, the highest-energy optical peak 
appears with a complex line-shape with a double-peak
structure for all the isotopes both at low and
high temperature. 
We made the hypothesis that this complex line-shape
is caused by the presence of two competing decay mechanisms; for example 
in the PdD case at the point ${\bf q} = [1/2,0,0] 2\pi/a$ we have a strong 
peak around 490~cm$^{-1}$, which is interestingly situated higher than the 
energy of the unperturbed SSCHA phonon eigenvalue (479.1~cm$^{-1}$). 
Another peak is present at a slightly lower energy, around 
450~cm$^{-1}$. We decomposed the contribution to 
$\Gamma^{\mathrm{ph-ph}}_{\mu}({\bf q},\omega) = - \mathfrak{Im} \Pi^{\mathcal{H}B}_{\mu}(\qq,\omega)$ at 
this point and the energy of the two peaks, over the BZ. We found that the 
higher peak is caused by the activation of a decay channel toward two phonons 
on the lower optical band: one at X, the other at the Gamma point. This decay 
mechanism is forbidden at the energy of the lower peak, where the favourite 
decay mechanism is toward one optical and one acoustical phonon.

Even if in INS experiments in non-stoichiometric 
palladium hydrides complex line-shapes  
with a large linewidth were 
observed~\cite{PhysRevLett.33.1297,PhysRevLett.57.2955},
larger than those calculated by us here, 
the origin of the structure has been attributed to 
the non-stoichiometry~\cite{PhysRevB.17.488}. Thus, we avoid the explicit 
comparison
with experiments as we are dealing with stoichiometric hydrides.


\section{Summary and conclusions}
\label{conclusions}

In this work we show how two different brand-new 
approaches can be combined to gain further insight
into the anharmonic properties of solids: the non-perturbative
stochastic self-consistent harmonic 
approximation~\cite{Errea13,Errea14},
which is valid to deal with strongly anharmonic crystals,
and the calculation of third-order anharmonic 
coefficients within DFPT and the ``$2n+1$'' theorem~\cite{Paulatto13}.
The combination of these two methods 
(i) yields the correct perturbative limit for
small anharmonicity and 
(ii) allows including self-energy corrections
to the variational SSCHA result.
Thus, the combination of the SSCHA with the
``$2n+1$'' theorem offers us the possibility
to calculate anharmonic phonon shifts and lifetimes
at any $\qq$ point in the BZ in a very
efficient way for crystals were perturbation
theory remains valid, avoiding the cumbersome   
calculation of fourth-order anharmonic coefficients.
Moreover, it also allows calculating phonon lifetimes
and INS spectra for those crystals where a non-perturbative
approach of anharmonicity is required.
Thus, this approach might be very appealing 
for calculating thermal conductivity in strongly anharmonic
crystals like thermoelectrics.

The combination of the SSCHA and the ``$2n+1$'' theorem
is applied to the strongly anharmonic palladium hydrides,
where the harmonic theory completely breaks down.
We show that despite being a superconducting metal 
the phonon lifetime is dominated by anharmonicity.
Even if the anharmonic phonon linewidth is huge and strongly temperature 
dependent, the phonon shift induced by the third-order is not predominant 
over the SSCHA correction of the phonon bands.
Nevertheless, the third-order broadening of the phonon modes is huge,
at the point that the simplistic model of well-distinct phonon modes, broadened 
to a finite-width Lorentzian function, cannot be applied anymore.
In fact, the INS spectra calculated show a complex and unexpected
structure for the optical modes with
satellite peaks, far from the standard Lorentzian
line-shape. We think that measuring 
these feature should be possible, albeit challenging, 
with INS techniques. Our calculations show that large anharmonicity
can induce very complex and unexpected INS spectra
that could be misleading for experimentalists. 

%
%

\section*{ACKNOWLEDGMENTS}

The authors acknowledge support from the Graphene Flagship,
from the French state funds managed by
the ANR within the Investissements d'Avenir programme under reference
ANR-13-IS10-0003-01,
and from the Spanish Ministry of Economy and Competitiveness (FIS2013-48286-C2-2-P).
I.E. would like to acknowledge financial support
also from the
Department of Education, Language Policy and Culture of the Basque
Government (Grant No. BFI-2011-65).
Computer facilities were provided by CINES, CCRT and IDRIS
(Project No. x2014091202), and 
the PRACE 9th Regular call (Project AESFT).


\appendix

\section{Fourier transforms and the $\phi_{\mu_1 \dots \mu_n}(\qq_1, 
\dots , \qq_n)$
         coefficients}
\label{appen-fourier}

Defining the Fourier transforms of the atomic displacements as
\beq
u_{s}^{\alpha}(\mathbf{R}) = \frac{1}{\sqrt{N}} \sum_{\qq} e^{-i 
\qq \mathbf{R}} u_{s}^{\alpha}(\qq) 
\label{fourier-u}
\eeq
and of the 
derivatives of the BO energy surface as
\beqn
\phi_{s_1 \dots s_n}^{\alpha_1 \dots \alpha_n} (\mathbf{R}_1, \dots, 
\mathbf{R}_n) & = &
\frac{1}{N^{n-1}} \sum_{\qq_1 \dots \qq_n} 
e^{i (\qq_1 \mathbf{R}_{1} + \dots + \qq_n \mathbf{R}_{n})} 
\nonumber \\
& \times & \phi_{s_1 \dots s_n}^{\alpha_1 \dots \alpha_n} (\qq_1, \dots, 
\qq_n),
\label{fourier-phi}
\eeqn
the BO potential can be written as
\beqn
V & = & V_0 + \sum_{n=2}^{\infty} \frac{N^{1-n/2}}{n!} \sum_{\substack{s_1 \dots 
s_n \\ \alpha_1 \dots \alpha_n \\ 
      \qq_1 \dots \mathbf{q_n}}}
    \phi_{s_1 \dots s_n}^{\alpha_1 \dots \alpha_n} (\qq_1, \dots, 
\qq_n) \nonumber \\
   & \times & u_{s_1}^{\alpha_1}(\qq_1) \dots  
u_{s_n}^{\alpha_n}(\qq_n).
\label{fourier-potential}
\eeqn
Making use of the change of variables in Eq.~\eqref{qmu} and the definition
\beqn
&& \phi_{\mu_1 \dots \mu_n}(\qq_1, \dots , \qq_n)  =   
\sum_{\substack{s_1 \dots s_n \\ \alpha_1 \dots \alpha_n}}
          \phi_{s_1 \dots s_n}^{\alpha_1 \dots \alpha_n} (\qq_1, \dots, 
\qq_n) \nonumber \\
    && \ \ \ \ \ \times   
         \sqrt{\frac{\hbar}{2 M_{s_1} \omega_{\mu_1} (\qq_1)}} \dots
         \sqrt{\frac{\hbar}{2 M_{s_n} \omega_{\mu_n} (\qq_n)}} \nonumber 
\\
    && \ \ \ \ \ \times 
         \epsilon_{\mu_1 s_1}^{\alpha_1} (\qq_1) \dots 
         \epsilon_{\mu_n s_n}^{\alpha_n} (\qq_n),       
\label{anhar-coeff}
\eeqn
it is easy to show that the ionic Hamiltonian can be written as in
Eq.~\eqref{potential-taylor-mode}.

\section{The perturbative limit of the SSCHA}
\label{appen-scha}

In perturbation theory the lowest order self-energy diagrams
involve third- and fourth-order terms. In order to understand 
the perturbative limit of the SSCHA we will therefore truncate 
the expansion in Eq.~\eqref{fourier-potential} at fourth order.
In that case the forces in Eqs. \eqref{gradient-eq} and
\eqref{gradient-phi} are
\beqn
&& f_s^{\alpha}(\qq) = - \sum_{s_2 \alpha_2} \phi_{s s_2}^{\alpha 
\alpha_2} (\qq)
                                 u_{s_2}^{\alpha_2}(\qq) \nonumber \\
&& \ \                       - \frac{1}{2\sqrt{N}}\sum_{\substack{s_2 s_3 \\ 
\alpha_2 \alpha_3 \\ \qq_2 \qq_3}}
                                \phi_{s s_2 s_3}^{\alpha \alpha_2 \alpha_3} (- 
\qq, \qq_2, \qq_3)
                                u_{s_2}^{\alpha_2}(\qq_2) 
u_{s_3}^{\alpha_3}(\qq_3)  
\nonumber \\
&& \ \         - \frac{1}{3! N}\sum_{\substack{s_2 s_3 s_4 \\ \alpha_2 \alpha_3 
\alpha_4 \\ \qq_2 \qq_3 \qq_4}}
               \phi_{s s_2 s_3 s_4}^{\alpha \alpha_2 \alpha_3 \alpha_4} (- 
\qq, \qq_2, \qq_3, \qq_4)
               u_{s_2}^{\alpha_2}(\qq_2) \nonumber \\
&& \ \ \ \ \ \ \ \ \ \ \ \ \ \ \ \times u_{s_3}^{\alpha_3}(\qq_3) 
u_{s_4}^{\alpha_4}(\qq_4)
\label{force-4th}
\eeqn
and
\beq
\mathfrak{f}_s^{\alpha}(\qq) = - \sum_{s_2 \alpha_2} \Phi_{s s_2}^{\alpha 
\alpha_2} (\qq)
                                 \mathfrak{u}_{s_2}^{\alpha_2}(\qq).
\label{force-sscha}
\eeq

\subsection{The tadpole diagram}

Making use of Eqs. \eqref{force-4th} and \eqref{force-sscha}, it is easy 
to show that in the perturbative limit Eq. \eqref{gradient-eq} reduces
to
\beqn
 \boldsymbol{\nabla}_{\mathbf{\mathfrak{R}}_{\mathrm{eq}}}\mathcal{F}_H[\mathcal{H}]   
  & = & \frac{1}{2\sqrt{N}}\sum_{\substack{s_2 s_3 \\ \alpha_2 \alpha_3 \\ \qq}}
                 \phi_{s s_2 s_3}^{\alpha \alpha_2 \alpha_3} (0, \qq, -\qq) \nonumber \\
  & \times &     \langle   u_{s_2}^{\alpha_2}(\qq) u_{s_3}^{\alpha_3}(-\qq) \rangle_{\mathcal{H}}.
\eeqn
Substituting $u_{s}^{\alpha}(\qq) = \mathfrak{u}_{s}^{\alpha}(\qq) - \sqrt{N} (
R_{{\rm eq}}^{s \alpha} - \mathfrak{R}_{{\rm eq}}^{s \alpha})$ in the equation above,
straightforwardly
\beqn
&& \boldsymbol{\nabla}_{\mathbf{\mathfrak{R}}_{\mathrm{eq}}}\mathcal{F}_H[\mathcal{H}]   
   =  \frac{1}{2\sqrt{N}}\sum_{\substack{s_2 s_3 \\ \alpha_2 \alpha_3 \\ \qq}}
                 \phi_{s s_2 s_3}^{\alpha \alpha_2 \alpha_3} (0, \qq, -\qq) \nonumber \\
  && \ \ \times   \Bigg[ \sum_{\mu} \frac{\hbar \varepsilon_{\mu s_2}^{\alpha_2} (\qq) 
                                      \varepsilon_{\mu s_3}^{\alpha_3 *} (\qq) }{
                          2 \Omega_{\mu}(\qq) \sqrt{M_{s_2}M_{s_3}}}
                          [2n_B(\Omega_{\mu}(\qq)) + 1] \nonumber \\
  && \ \ \ \ \ + N (R_{{\rm eq}}^{s_2 \alpha_2} - \mathfrak{R}_{{\rm eq}}^{s_2 \alpha_2})
                 (R_{{\rm eq}}^{s_3 \alpha_3} - \mathfrak{R}_{{\rm eq}}^{s_3 \alpha_3}) \Bigg].
\label{pert-tadpole}
\eeqn
Equalizing the gradient to zero the renormalized positions are obtained. If we
substitute the $\Omega_{\mu}(\qq)$ phonon frequencies and $\varepsilon_{\mu s}^{\alpha} (\qq)$
polarization vectors by the harmonic phonon frequencies obtained for a 
given value of the internal coordinates, comparing Eq. \eqref{pert-tadpole} 
with Eq. \eqref{pi-t} it is easy to note that for the renormalized
internal coordinates the T$_\mathrm{O}$ self-energy vanishes. 
This is the result of the QHA. As shown in Eq. \eqref{pert-tadpole}, 
the lowest-order correction given by the SSCHA is thus
the QHA.  

\subsection{The loop diagram}

In the perturbative limit for a fixed cell 
geometry~\cite{PhysRevB.68.220509,Calandra07}
the equilibrium positions do not change and polarization
vectors are kept at the harmonic level. Thus, 
$\mathfrak{u}_{s}^{\alpha}(\qq) = u_{s}^{\alpha}(\qq)$
and $\varepsilon_{\mu s}^{\alpha} (\qq) = \epsilon_{\mu s}^{\alpha} 
(\qq)$.
Within this assumptions, it is straightforward to show that
Eq.~\eqref{gradient-phi} becomes
\beqn
&& \boldsymbol{\nabla}_{\Phi(\qq)} \mathcal{F}_H[\mathcal{H}] =
   \sum_{\mu} 
     \Bigg(
       \left[ \omega^2_{\mu}(\qq) - \Omega^2_{\mu}(\qq) \right] 
\nonumber \\ 
&& \ \ \ \ \ \
     + \frac{1}{2N} \sum_{\qq' \mu'} \sum_{\substack{s_1 s_2 s_3 s_4 \\ 
\alpha_1 \alpha_2 \alpha_3 \alpha_4}}
       \phi_{s_1 s_2 s_3 s_4}^{\alpha_1 \alpha_2 \alpha_3 \alpha_4} 
(-\qq, \qq, \qq', -\qq')
\nonumber \\ 
&& \ \ \ \ \ \
     \times \frac{ \epsilon_{\mu s_1}^{\alpha_1 *} (\qq ) 
                  \epsilon_{\mu s_2}^{\alpha_2  } (\qq ) 
                  \epsilon_{\mu's_3}^{\alpha_3  } (\qq') 
                  \epsilon_{\mu's_4}^{\alpha_4 *} (\qq') }{ 
                          \sqrt{M_{s_1} M_{s_2} M_{s_3} M_{s_4}}} 
\mathfrak{a}^2_{\mu'}(\qq')
                 \Bigg) 
\nonumber \\ 
&& \ \ \ \ \times
     \mathfrak{a}^2_{\mu}(\qq) \boldsymbol{\nabla}_{\Phi(\qq)} \ln 
\mathfrak{a}_{\mu}(\qq).
\label{gradient-phi-4th}
\eeqn
Therefore, the gradient will vanish if the self-consistent
\beqn
\Omega^2_{\mu}(\qq) & = & \omega^2_{\mu}(\qq) + 
        \frac{1}{2N} \sum_{\substack{ \qq' \mu' \\ s_1 s_2 s_3 s_4 \\ 
\alpha_1 \alpha_2 \alpha_3 \alpha_4}}
       \phi_{s_1 s_2 s_3 s_4}^{\alpha_1 \alpha_2 \alpha_3 \alpha_4} 
(-\qq, \qq, \qq', -\qq')
\nonumber \\ 
& \times &
       \frac{ \epsilon_{\mu s_1}^{\alpha_1 *} (\qq ) 
                  \epsilon_{\mu s_2}^{\alpha_2  } (\qq ) 
                  \epsilon_{\mu's_3}^{\alpha_3  } (\qq') 
                  \epsilon_{\mu's_4}^{\alpha_4 *} (\qq') }{ 
                          \sqrt{M_{s_1} M_{s_2} M_{s_3} M_{s_4}}} 
                           \mathfrak{a}^2_{\mu'}(\qq')
\label{self-consistent-eq}
\eeqn
equation is fulfilled. Eq.~\eqref{self-consistent-eq}
is the same equation that was used in Ref.~\onlinecite{PhysRevLett.106.165501}
to apply the SSCHA to simple cubic calcium.
At lowest order we can assume that 
$\mathfrak{a}^2_{\mu'}(\qq') \sim \hbar \coth (\beta \hbar \omega_{\mu'} 
(\qq') / 2) / 
                                         (2 \omega_{\mu'} (\qq'))$
so that the equation above can be rewritten as
\beqn
\Omega^2_{\mu}(\qq) & = & \omega^2_{\mu}(\qq) + 
        \frac{\omega_{\mu} (\qq)}{\hbar N} \sum_{\qq' \mu' }
       \phi_{\mu \mu \mu' \mu'} (-\qq, \qq, \qq', 
-\qq')
\nonumber \\ 
& \times &
       [2n_B(\omega_{\mu'}(\qq')) + 1] 
\label{self-consistent-eq2}
\eeqn
or, in terms of the loop self-energy of Eq.~\eqref{pi-l},
\beq
\Omega^2_{\mu}(\qq) = \omega^2_{\mu}(\qq) + \frac{2 \omega_{\mu} 
(\qq)}{\hbar}
                             \Pi^{L}_{\mu}(\qq,\omega).       
\label{self-consistent-eq3}
\eeq
According to Eq.~\eqref{self-consistent-eq3}, the renormalized frequency 
given by the perturbative limit of the SSCHA is exactly at the peak
of the of the phonon propagator as long as the self-energy is given 
by the loop diagram~\cite{Mahan}. Thus,
in the perturbative limit the SCHA accounts exactly for the
shift given by the loop diagram.

\end{document}